\documentclass[%
 reprint,
 amsmath,amssymb,
 aps,
 prd,
]{revtex4-2}

\usepackage{graphicx}
\usepackage{float}
\usepackage{dcolumn}
\usepackage{bm}
\usepackage{multirow}
\usepackage{subcaption}

\begin{document}

\preprint{APS/123-QED}

\title{Search for the Higgs boson pair production in the $\bm{b\bar{b}\mu^+\mu^-}$ final state at the LHC}

\author{Botao Guo}
 \email{botao.guo@stu.pku.edu.cn}
\author{Xiaohu Sun}
 \email{Xiaohu.Sun@pku.edu.cn, corresponding author}
\author{Licheng Zhang}%
\author{Zhe Li}
\author{Yong Ban}
\affiliation{%
Department of Physics and State Key Laboratory of Nuclear Physics and Technology, Peking University, Beijing 100871, China
}%




\date{\today}

\begin{abstract}
The Higgs boson pair production via gluon-gluon fusion and vector boson fusion in the $b\bar{b}\mu^+\mu^-$ final state at the LHC is studied to probe the Higgs self-coupling $\kappa_\lambda$ and the four-boson HHVV coupling $\kappa_{\rm{2V}}$ for the first time. A cut-based analysis and a machine learning analysis using boosted decision trees are performed with categorizations and optimizations depending on the variations of these couplings. The expected sensitivities are extracted with different integrated luminosities assumed up to the full HL-LHC runs. The expected upper limit at 95\% confidence level on the HH production is calculated as 47 (28) times the Standard Model cross-section using the cut-based method (boosted decision trees) for the gluon-gluon fusion production, and 928 for the vector boson fusion production, assuming an integrated luminosity of 3000 fb$^{-1}$. The expected constraints on the couplings at 95\% confidence level are calculated to be $-13.8< \kappa_{\lambda}< 19.1$ ($-10.0< \kappa_{\lambda}< 15.5$) and $-3.4< \kappa_{\rm{2V}}< 5.5$ using the cut-based method (boosted decision trees), respectively, assuming an integrated luminosity of 3000 fb$^{-1}$.
\end{abstract}

\maketitle


\section{\label{sec:level1}Introduction}

Since the discovery of the Higgs boson by the ATLAS and CMS collaborations \cite{ATLAS:2012yve,CMS:2012qbp}, the measurements of its properties are of high priorities in order to understand the Brout–Englert–Higgs mechanism~\cite{Englert:1964et,Higgs:1964ia,Higgs:1964pj,Guralnik:1964eu,Higgs:1966ev,Kibble:1967sv} in the Standard Model (SM). The self-interaction of the Higgs boson scaled by the relative self-coupling $\kappa_\lambda = \lambda / \lambda_{\text{SM}}$ is fundamental in the determination of the shape of the Higgs potential. The Higgs boson pair production (HH) is the only accessible mode in the direct probe of the Higgs self-coupling at the Large Hadron Collider (LHC) and the high-luminosity LHC (HL-LHC). Given its low rates at the LHC, any enhancement of the HH production can also indicate physics beyond the Standard Model (BSM)~\cite{Grojean:2004xa,Cao:2013si,Gouzevitch:2013qca,Gupta:2013zza,Han:2013sga,Nishiwaki:2013cma,Goertz:2014qta,Hespel:2014sla,Cao:2014kya,Azatov:2015oxa,Carena:2015moc,Grober:2015cwa,Wu:2015nba,He:2015spf,Carvalho:2015ttv,Zhang:2015mnh,Huang:2015tdv,Nakamura:2017irk,DiLuzio:2017tfn,Huang:2017nnw,Buchalla:2018yce,Borowka:2018pxx,Chang:2019vez,Blanke:2019hpe,Li:2019tfd,Capozi:2019xsi,Alves:2019igs,Kozaczuk:2019pet,Barducci:2019xkq,Huang:2019bcs,Cheung:2020xij,Cao:2015oaa,Cao:2016zob,Li:2019uyy,Lu:2015qqa,Ren:2017jbg}.

\begin{figure}[hbt!]
    \centering
    \includegraphics[width=0.45\textwidth]{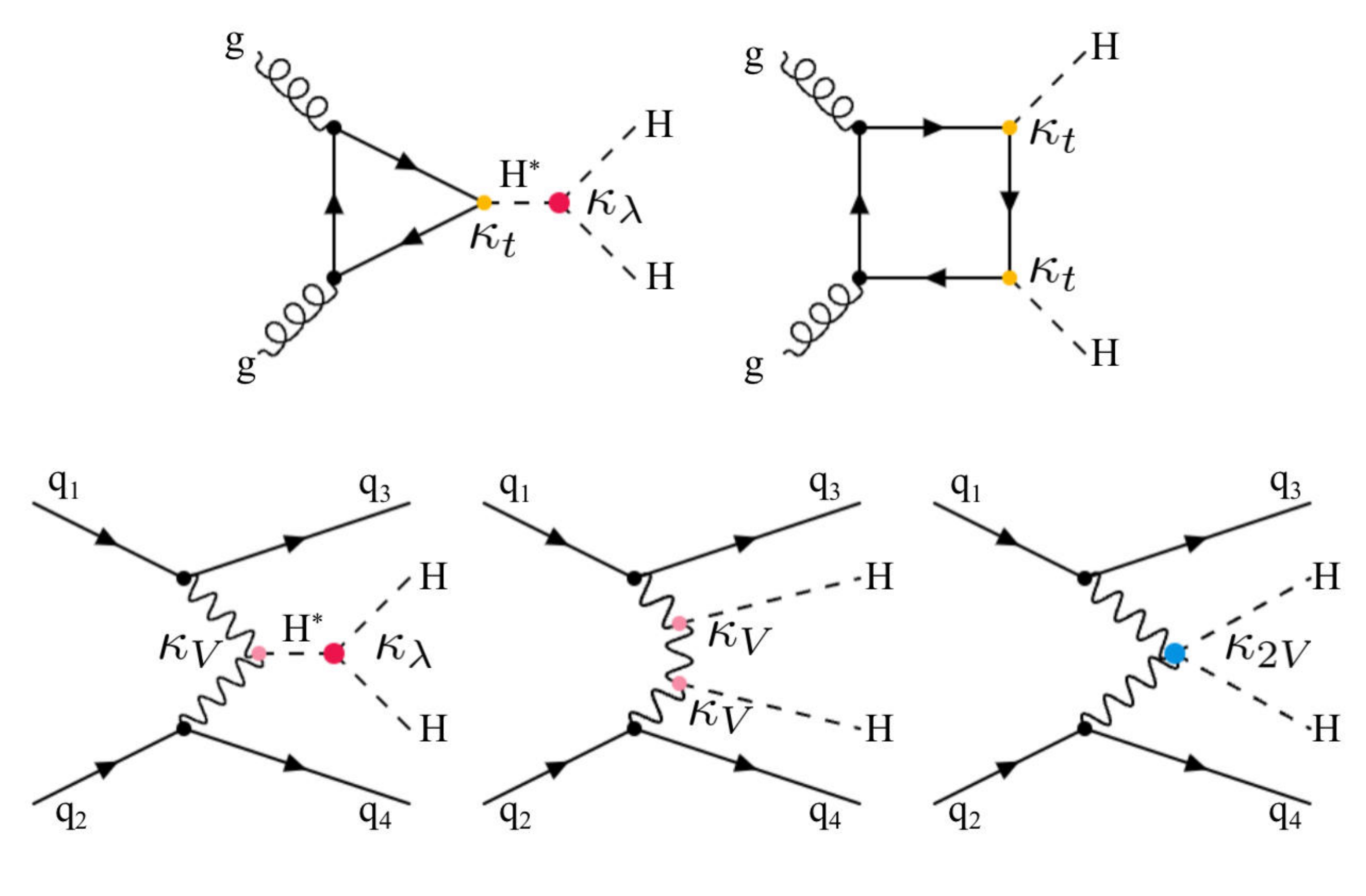}
    \caption{Leading-order diagrams contributing to the Higgs pair production: gluon-gluon fusion (ggF) and vector-boson fusion (VBF). The ggF mode contains the trilinear Higg
    self-coupling $\kappa_{\lambda}$ and the top quark Yukawa coupling $\kappa_{\rm{t}}$. The VBF mode contains the four-boson HHVV coupling $\kappa_{\text{2V}}$, the HVV coupling $\kappa_{\text{V}}$ and the trilinear Higgs self-coupling $\kappa_{\lambda}$.}
    \label{fig:figure_1}
\end{figure}

In proton-proton collisions, the largest HH production mode is gluon-gluon fusion (ggF) with a cross-section of 31.05$^{+6\%}_{-23\%}$(scale+$m_{\rm{t}}$)$\pm 3.0\%$(PDF+$\alpha_{\rm{S}}$) fb calculated at next-to-next-to-leading order (NNLO) in QCD with top quark mass effects (FT$_{\rm{approx}}$)~\cite{Grazzini:2018bsd} at centre-of-mass energy ($\sqrt{s}$) of 13 TeV. This is followed by vector boson fusion (VBF) with a cross-section of 1.726$^{+0.03\%}_{-0.04\%}$(scale)$\pm 2.1\%$(PDF+$\alpha_{\rm{S}}$) fb calculated at next-to-next-to-next-to-leading order (N$^3$LO) in QCD~\cite{Dreyer:2018qbw}. The ggF mode provides a direct probe of $\kappa_\lambda$ at the leading order (LO), while the VBF mode can additionally open a window to the four-boson HHVV coupling $\kappa_{\text{2V}}$,  as shown in FIG.~\ref{fig:figure_1}. Other production modes such as VHH, ttHH and tjHH with much smaller cross-sections~\cite{Baglio:2012np,Frederix:2014hta} are not discussed here.

Given the two Higgs bosons simultaneously produced in the process, there is a variety in its final states to explore. Recent experimental results performed by ATLAS and CMS cover the decay channels of $b\bar{b}b\bar{b}$ \cite{CMS:2022cpr,ATLAS:2018rnh,CMS:2018sxu,CMS:2018vjd}, $b\bar{b}\gamma\gamma$ \cite{ATLAS:2021ifb,CMS:2020tkr,ATLAS:2018dpp,CMS:2018tla}, $b\bar{b}\tau^{+}\tau^{-}$ \cite{CMS:2022hgz,CMS:2021yci,ATLAS:2018uni}, $b\bar{b}\rm{ZZ^{*}}$ \cite{CMS:2020jeo,CMS:2022omp}, $b\bar{b}\rm{WW^{*}}$ \cite{CMS:2017rpp,CMS:2019noi,ATLAS:2018fpd}, $\rm{WW^{*}WW^{*}}$ \cite{ATLAS:2018ili,CMS:2022kdx},$\rm{WW^{*}}\tau^{+}\tau^{-}$ \cite{CMS:2022kdx}, $\rm{WW^{*}}\gamma\gamma$ \cite{ATLAS:2018hqk}, and $\tau^{+}\tau^{-}\tau^{+}\tau^{-}$ \cite{CMS:2022kdx}. In terms of the expected sensitivity, the upper limits from the leading decay channels reach around 4 times the SM prediction on the ggF HH cross-section, while the combined results start to get close to 2 times~\cite{CMS:2022dw,ATLAS-CONF-2022-050,ATLAS:2019qdc,ATLAS:2015sxd,CMS:2018ipl}.
By now, the searches of the HH production has covered the Higgs decays to bosons and third-generation fermions, mostly benefiting from the large branching ratios, while they have not fully explored the decays involving the second-generation fermions, such as $\text{HH} \to b\bar{b}\mu^+\mu^-$. This is a rare decay channel but can benefit from the excellent mass resolution in the di-muon invariant mass $m_{\mu\mu}$ at a level of 1-2 GeV \cite{CMS:2020xwi}, similar to $m_{\gamma\gamma}$, at the LHC. Moreover, CMS~\cite{CMS:2020xwi} has claimed the evidence of $\text{H}\to\mu\mu$ and ATLAS~\cite{ATLAS:2020fzp} also found large excess in this final state. Experimentally this decay channel becomes feasible.

We present a comprehensive study of the HH searches in $\text{HH}\to b\bar{b}\mu^+\mu^-$, focusing on not only ggF but also VBF production modes, with a dependence on the Higgs self-coupling $\kappa_\lambda$ and the HHVV-coupling $\kappa_{\text{2V}}$. In measuring $\kappa_\lambda$ and $\kappa_{\text{2V}}$, other couplings, such as $\kappa_\text{V}$ and $\kappa_\text{t}$ that directly enter the leading HH diagrams, are all set to their SM values, following many of the recent ATLAS and CMS HH analyses. The measurements of $\kappa_\text{V}$ and $\kappa_\text{t}$ mostly rely on single Higgs processes and thus are not discussed in the scope of this paper. Both the cut-based and boosted decision trees (BDT) methods are applied in the optimization of event selections. The final fits are performed on the kinematic shapes including the di-muon invariant mass $m_{\mu\mu}$ together with the di-bjet invariant mass $m_{bb}$. For a comparison to the existing literature, Ref.~\cite{Baur:2003gp} briefly discussed $\text{HH}\to b\bar{b}\mu^+\mu^-$ with a cut-based method using counting experiments, only targeting at the SM ggF HH production. Ref.~\cite{Adhikary:2020fqf} considered $\text{HH}\to b\bar{b}\mu^+\mu^-$ with a multivariate analysis using the BDT algorithm at the high-energy LHC (HE-LHC).


This paper is structured as follows. Section~\ref{sec:samples} introduces the simulated signal and background samples. Section~\ref{sec:evt} describes the analysis strategies, event categorizations and selections. Section~\ref{sec:bdt} focuses on the BDT analysis. Finally, Section~\ref{sec:res} reports the results and Section~\ref{sec:sum} summarizes the conclusions.

\section{Signal and background samples}
\label{sec:samples}

For the signal Monte Carlo (MC) samples, the ggF HH processes are generated at next-to-leading order (NLO) in QCD using POWHEG-BOX-V2~\cite{Heinrich:2017kxx,Heinrich:2019bkc}, while the VBF HH processes are generated at LO in QCD using MG5\_aMC@NLO V2.6.5 \cite{Alwall:2014hca}. In total, 7 ggF HH MC samples are generated with different $\kappa_{\lambda}$ values, $\kappa_\lambda$ = -5, 0, 1, 2.4, 5, 10 and 20, and 7 VBF HH MC samples are generated with different $\kappa_{2\rm{V}}$ values, $\kappa_{\text{2V}}$ = -10, -5, 0, 1, 2, 5 and 10. Other coupling values used in the scan later are from the combination of the generated samples, since the differential cross-section is scaled by a second-order polynomial of $\kappa_\lambda$ and $\kappa_{2\rm{V}}$ in the LO electroweak precision. This approach of linearly combining samples from different couplings follows the treatment in Ref.~\cite{CMS:2022cpr}.

\begin{figure}[hbt!]
    \centering
    \includegraphics[width=0.45\textwidth]{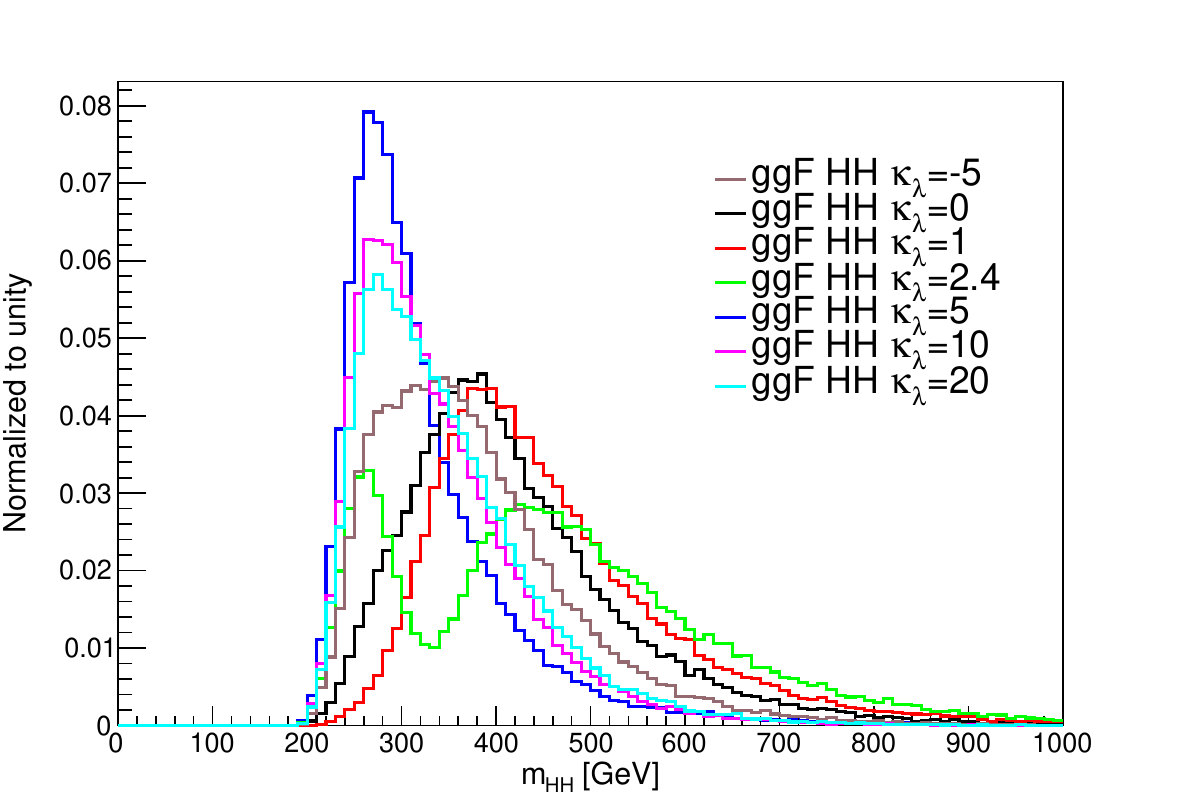}
    \caption{The $m_{\rm{HH}}$ distribution of ggF HH signals with different $\kappa_{\lambda}$ values.}
    \label{fig:figure_2}
\end{figure}

\begin{figure}[hbt!]
    \centering
    \includegraphics[width=0.45\textwidth]{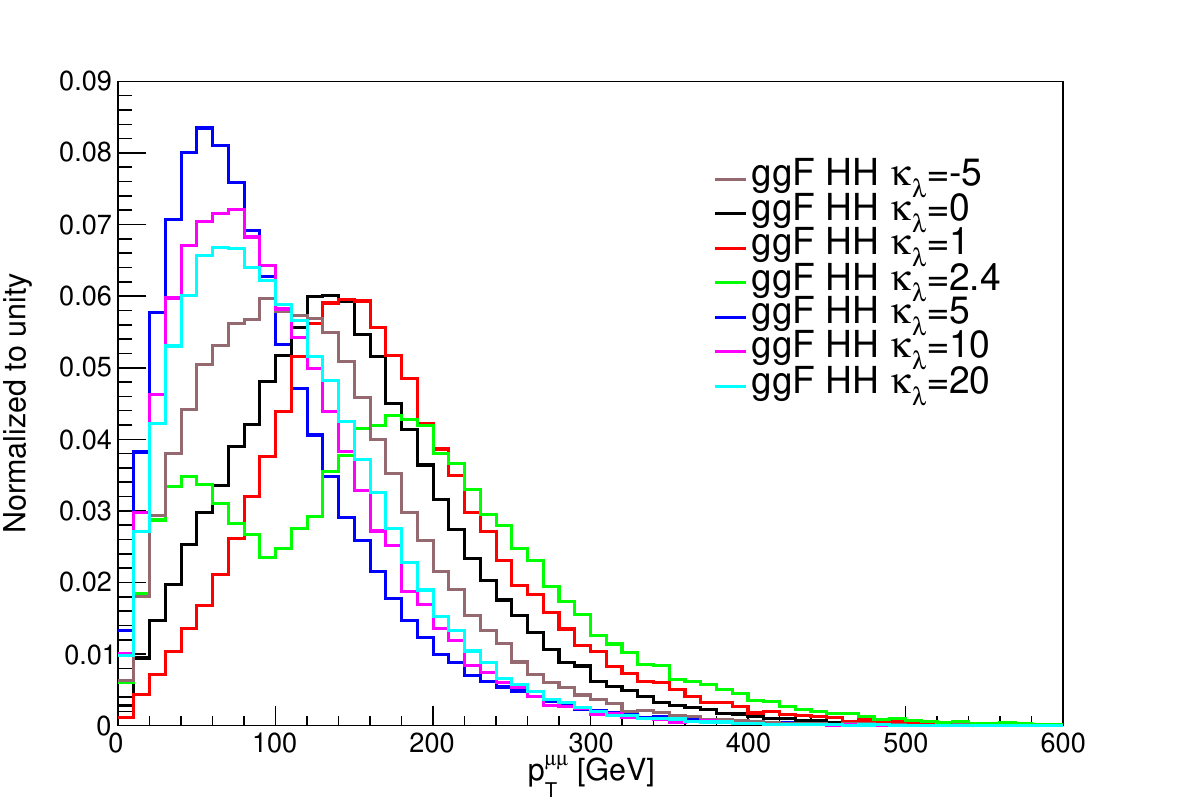}
    \caption{The $p_{\rm{T}}^{\mu\mu}$ distribution of ggF HH signals with different $\kappa_{\lambda}$ values.}
    \label{fig:figure_3}
\end{figure}

\begin{figure}[hbt!]
    \centering
    \includegraphics[width=0.45\textwidth]{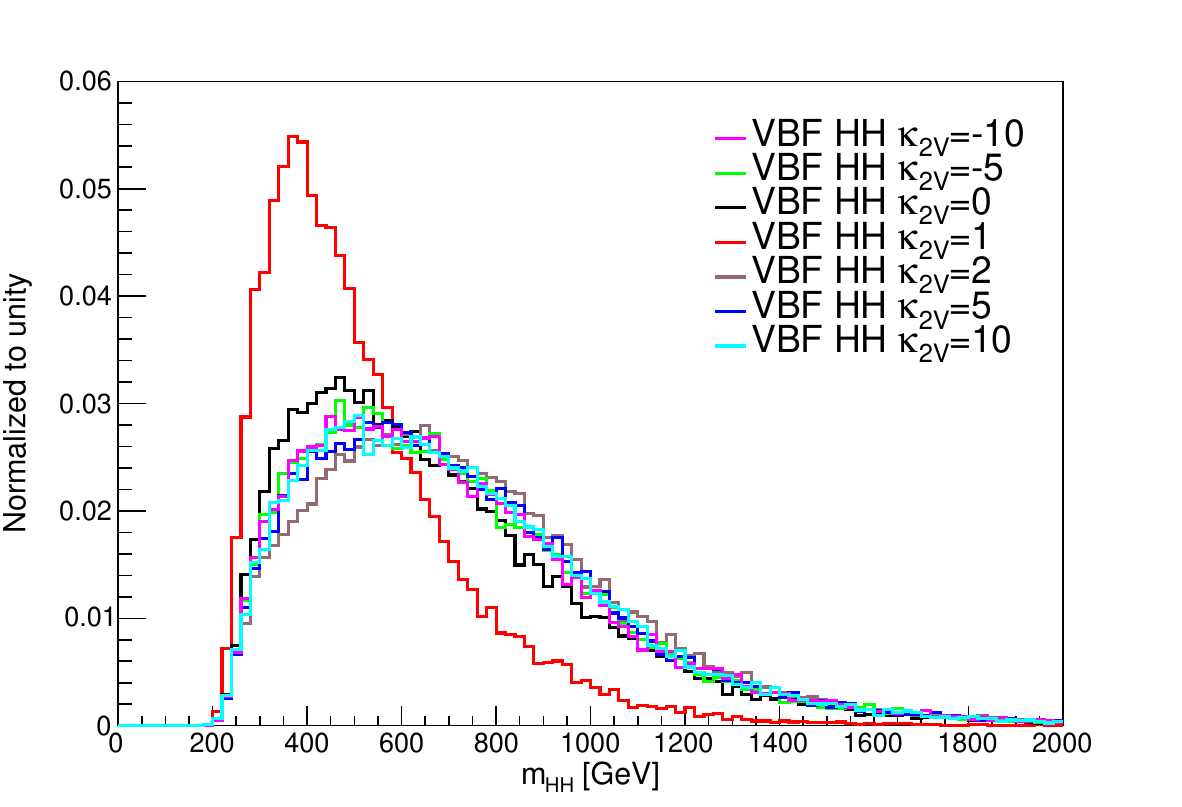}
    \caption{The $m_{\rm{HH}}$ distribution of VBF HH signals with different $\kappa_{2\rm{V}}$ values.}
    \label{fig:figure_4}
\end{figure}

\begin{figure}[hbt!]
    \centering
    \includegraphics[width=0.45\textwidth]{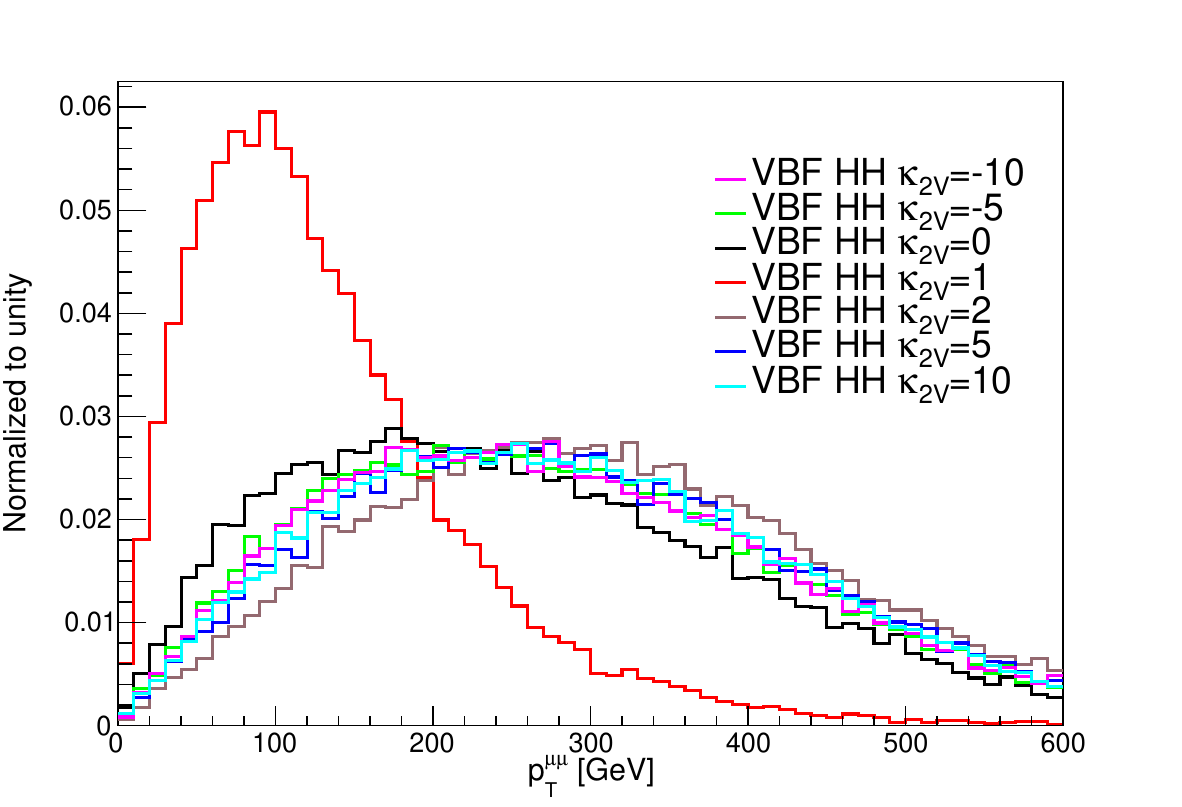}
    \caption{The $p_{\rm{T}}^{\mu\mu}$ distribution of VBF HH signals with different $\kappa_{2\rm{V}}$ values.}
    \label{fig:figure_5}
\end{figure}

The relevant couplings modify not only the total rates but also the kinematics significantly, as shown in FIG.~\ref{fig:figure_2},~\ref{fig:figure_3},~\ref{fig:figure_4} and~\ref{fig:figure_5}, where $m_{\rm{HH}}$ is the reconstructed invariant mass of $b\bar{b}\mu^+\mu^-$, and $p_{\rm{T}}^{\mu\mu}$ is the transverse momentum of the $\mu\mu$ system. In these figures, all distributions are normalized to unity in order to have a direct comparison on the shapes. For ggF HH process, in FIG.~\ref{fig:figure_2}, distributions of the invariant mass $m_{\rm{HH}}$ of the Higgs boson pair system are displayed for different values of $\kappa_{\lambda}$. They exhibit a characteristic dip at $m_{\rm{HH}}$ $\sim$ 350 GeV for $\kappa_{\lambda}$ $\sim$ 2.4~\cite{Heinrich:2020ytv}. This value of the trilinear Higgs self-coupling corresponds to a maximal destructive interference between the triangle and box diagrams in FIG.~\ref{fig:figure_1}. For $\kappa_{\lambda}$ = 1, the maximal destructive interference happens at the HH production threshold and therefore does not introduce a visible dip in the distribution. For larger $\kappa_{\lambda}$ values, the triangle diagram starts to dominate resulting in a softer spectrum. The similar interference structure is manifested in the distribution of $p_{\rm{T}}^{\mu\mu}$ in FIG.~\ref{fig:figure_3}. In the VBF process, the distributions tend to be harder for the BSM cases with $\kappa_{\text{2V}}$ deviating from 1, as shown in FIG.~\ref{fig:figure_4} and FIG.~\ref{fig:figure_5}.




Among the background processes, the Drell-Yan (DY) process is dominant in general. The DY samples are generated with at least 1 b-quark and up to 4 quarks associated, using MG5\_aMC@NLO V2.6.5. The generated DY processes include $llb$, $llbj$, $llbjj$ and $llbjjj$, where $l$ stands for muon, $b$ for bottom quark and $j$ for all quarks except top quark. To enhance the MC statistics in the signal enriched phase space, the samples are generated with $m_{ll}$ slices of [100, 150], [150, 200] and [200, $+\infty$] GeV. A k-factor of 1.23 \cite{LHC-k-factor} is used for the NLO QCD correction in the DY process. No k-factors are applied for other background processes, given the fact that DY totally dominates the background contribution. The second leading background is top quark pair production $t\bar{t}$. The samples are generated at LO in QCD using MG5\_aMC@NLO V2.6.5. Single Higgs processes that have a Higgs boson decaying to a pair of muons can also enter the signal regions. These samples are generated in the production modes of gluon-gluon fusion (ggH), vector boson fusion (VBFH), ZH, $\rm{t\bar{t}H}$ and $\rm{b\bar{b}H}$, using MG5\_aMC@NLO V2.6.5.

For all signal and background samples, the decay, parton shower, hadronization and underlying events are modelled by PYTHIA 8.306~\cite{Bierlich:2022pfr}. No pileup is considered. To emulate the detector effects, DELPHES 3.5.0~\cite{deFavereau:2013fsa} is used and the default CMS configuration card is applied. The jets are reconstructed using the anti-$k_{t}$ clustering algorithm~\cite{Cacciari:2008gp,Cacciari:2011ma} with a radius parameter $R = 0.5$ following the CMS defaults. All the MC samples are summarized in TAB.~\ref{table:1} where the DY and $t\bar{t}$ cross-section contains the branching ratios down to the final state with di-muon, while the single and double Higgs processes do not contain any branching ratio.

\begin{table}[h!]
\renewcommand{\arraystretch}{1.5}
\centering
\begin{tabular}{ l c c c } 
\hline
\hline
Process\hspace{0.15cm} & $m_{l^{+}l^{-} }[\rm{GeV}]$\hspace{0.15cm} & $\sigma[\rm{fb}]$\hspace{0.15cm} & $N^{\rm{gen}}_{\rm{events}}(\times10^{6})$\hspace{0.15cm} \\
\hline
Drell-Yan & $[100,150]$ & 5481 & 9.98 \\ 
Drell-Yan & $[150,200]$ & 384 & 10.0 \\ 
Drell-Yan & $[200,+\infty]$ & 201 & 1.0 \\ 
$t\bar{t}$ & | & 4864 & 2.0 \\
ggH & | & 48580 & 1.0 \\
VBFH & | & 3782 & 1.0 \\
ZH & | & 883.9 & 1.0 \\
ttH & | & 507.1 & 1.0 \\
bbH & | & 488 & 1.0 \\
\hline 
ggF signal & & & \\
\hline
$\kappa_{\lambda}=-5$ & | & 599 & 0.55 \\
$\kappa_{\lambda}=0$ & | & 70 & 0.55 \\
$\kappa_{\lambda}=1$ & | & 31 & 0.55  \\
$\kappa_{\lambda}=2.4$ & | & 13 & 0.55  \\
$\kappa_{\lambda}=5$ & | & 95 & 0.55  \\
$\kappa_{\lambda}=10$ & | & 672 & 0.55  \\
$\kappa_{\lambda}=20$ & | & 3486 & 0.55  \\
\hline
VBF signal & & & \\
\hline
$\kappa_{2\rm{V}}=-10$ & | & 2365 & 0.50  \\
$\kappa_{2\rm{V}}=-5$ & | & 722 & 0.50  \\
$\kappa_{2\rm{V}}=0$ & | & 27 & 0.50  \\
$\kappa_{2\rm{V}}=1$ & | & 1.73 & 0.50  \\
$\kappa_{2\rm{V}}=2$ & | & 14.2 & 0.50  \\
$\kappa_{2\rm{V}}=5$ & | & 279 & 0.50  \\
$\kappa_{2\rm{V}}=10$ & | & 1479 & 0.50  \\
\hline
\hline
\end{tabular}
\caption{Summary of Monte Carlo samples.The first column gives the names of the processes. The second column provides the slicing on $m_{l^{+}l^{-}}$ whenever it is applicable. The third column lists the cross-section. The last column presents the number of events in the generated MC samples. The DY and $t\bar{t}$ cross-section contains the branching ratios down to the final state with di-muon, while the single and double Higgs processes do not contain any branching ratio.}
\label{table:1}
\end{table}

\section{Cut-based Analysis}
\label{sec:evt}

\subsection{Analysis Strategy}
\label{subsec:as}

Two analysis strategies are adopted. One using sequential cuts inspired by recent ATLAS and CMS analyses~\cite{ATLAS:2021ifb,CMS:2020tkr} serves as a baseline strategy to understand the basic kinematics and get a conservative estimation of the sensitivity, while the other applies the boosted decision trees (BDT) to seek further improvements in the sensitivity. The latter one is only studied for the ggF HH as there is insufficient statistics in the VBF enriched regions.

\subsection{Object and Basic Event Selection}

Basic acceptance requirements following a CMS-like detector are applied in the physics object selections. For muon candidates, they must satisfy the requirements of $p_{\rm{T}}>20\ \rm{GeV}$ and $|\eta|<2.4$ following~\cite{CMS:2020xwi}. Jets are reconstructed using the anti-$k_{t}$ clustering algorithm with a radius parameter $R = 0.5$. All jet candidates have to pass $p_{\rm{T}}>20\ \rm{GeV}$ within $|\eta|<4.7$, while the b-jets have to be within $|\eta|<2.4$ limited by the tracker geometry. The b-jets are tagged with an parameterized efficiency depending on $p_{\rm{T}}$ and $\eta$ mimicking the CMS scenario.

Each event is required to contain at least two opposite charged muons. When more than two muon candidates are found, the muon pair with the highest transverse momentum $p_{\rm{T}}^{\mu\mu}$ is chosen to reconstruct the Higgs boson candidate. Moreover, all the events have to sit in $100<m_{\mu\mu}<180\ \rm{GeV}$. Next, events are required to contain at least two b-jets. In case of more than two b-jets, the Higgs boson candidate is reconstructed from the two jets with the highest transverse momentum $p_{\rm{T}}^{bb}$. In the end, the invariant mass of two b-jets is required to be $70<m_{bb}<190\ \rm{GeV}$. Both the choices of the di-muon and di-bjet mass range are inspired by Ref.~\cite{CMS:2020tkr}. In the low $m_{\mu\mu}$ or $m_{bb}$ end, the statistics is much higher than the high end. Thus, the intervals are skewed towards higher mass ranges.

Besides the two b-jets, the signal events could have more jets in the VBF HH production. It is featured by the presence of two additional energetic jets (VBF jets), corresponding to two quarks from each of the colliding protons scattered away from the beam line. These VBF jets are expected to have a large spacial separation, which results in a large di-jet invariant mass, $m_{jj}^{\rm{VBF}}$. The jet pair with the highest $m_{jj}^{\rm{VBF}}$ in an event is selected as the two VBF jets.

\subsection{Event Categorization and Background Rejection}
\label{subsec:ec-br}

The events are firstly grouped by the signal production modes ggF and VBF. In each group, events are then categorized according to the couplings variations. After categorization, the sequential cuts are optimised in the cut-based analysis, and the training of BDT is performed in the machine learning analysis, both for suppressing the corresponding background in each category. Eventually, the fits are performed on the combined $m_{\mu\mu}$ and $m_{bb}$ distributions.

\begin{table}[h!]
\renewcommand{\arraystretch}{1.5}
\centering
\begin{tabular}{ c c c } 
\hline
\hline
Category\hspace{0.2cm} & $m_{jj}^{\rm{VBF}}$(GeV)\hspace{0.3cm} & $m_{\rm{HH}}^{\rm{corr}}$(GeV)\hspace{0.3cm} \\
\hline
ggF SM & $<880$ & $>400$ \\ 
ggF BSM & $<880$ & $<400$ \\
VBF SM & $>880$ & $<680$ \\
VBF BSM & $>880$ & $>680$ \\
\hline
\hline
\end{tabular}
\caption{Summary of the event categorization.}
\label{table:2}
\end{table}

\begin{figure}[hbt!]
    \centering
    \includegraphics[width=0.48\textwidth]{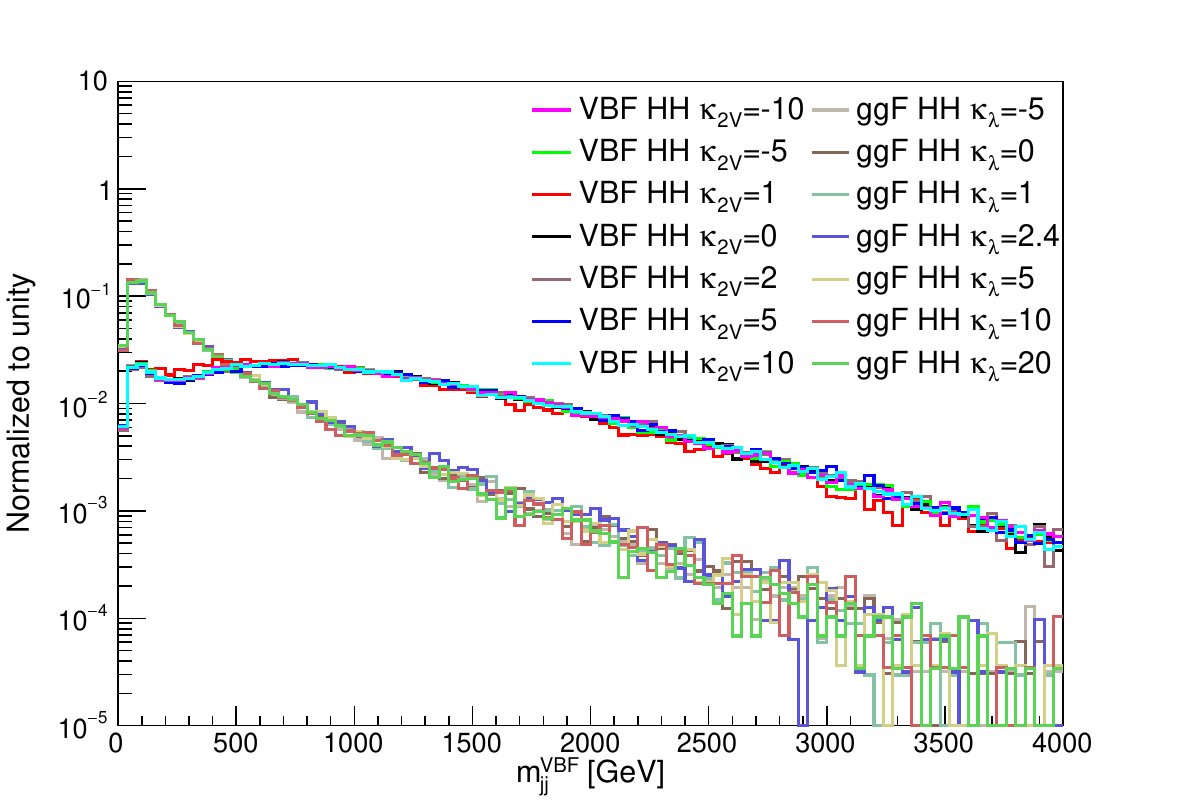}
    \caption{The $m_{jj}^{\rm{VBF}}$ distribution of all the signals.}
    \label{fig:figure_6}
\end{figure}

In the separation of ggF and VBF modes, events that do not have two or more additional jets enter the ggF categories directly. With two or more additional jets, events can be categorized by the invariant mass of the two VBF jets $m_{jj}^{\rm{VBF}}$, as shown in FIG.~\ref{fig:figure_6}. The VBF events clearly have a much harder spectrum of $m_{jj}^{\rm{VBF}}$ with respect to the ggF ones due to the VBF jets that fly out in a large spacial separation, and this feature does not depend either of the couplings $\kappa_\lambda$ and $\kappa_{\text{2V}}$. A scan on $m_{jj}^{\rm{VBF}}$ is performed to maximize the separation of the two production modes. This determines the threshold $m_{jj}^{\rm{VBF}} = 880$ GeV which the ggF category is defined below and VBF is defined above. Other variables, such as the difference in pseudorapidity between the two VBF jets $|\Delta\eta_{jj}^{\rm{VBF}}|$, are also tried, but do not show any significant improvement once $m_{jj}^{\rm{VBF}}$ is used.

\begin{figure}[hbt!]
    \centering
    \includegraphics[width=0.48\textwidth]{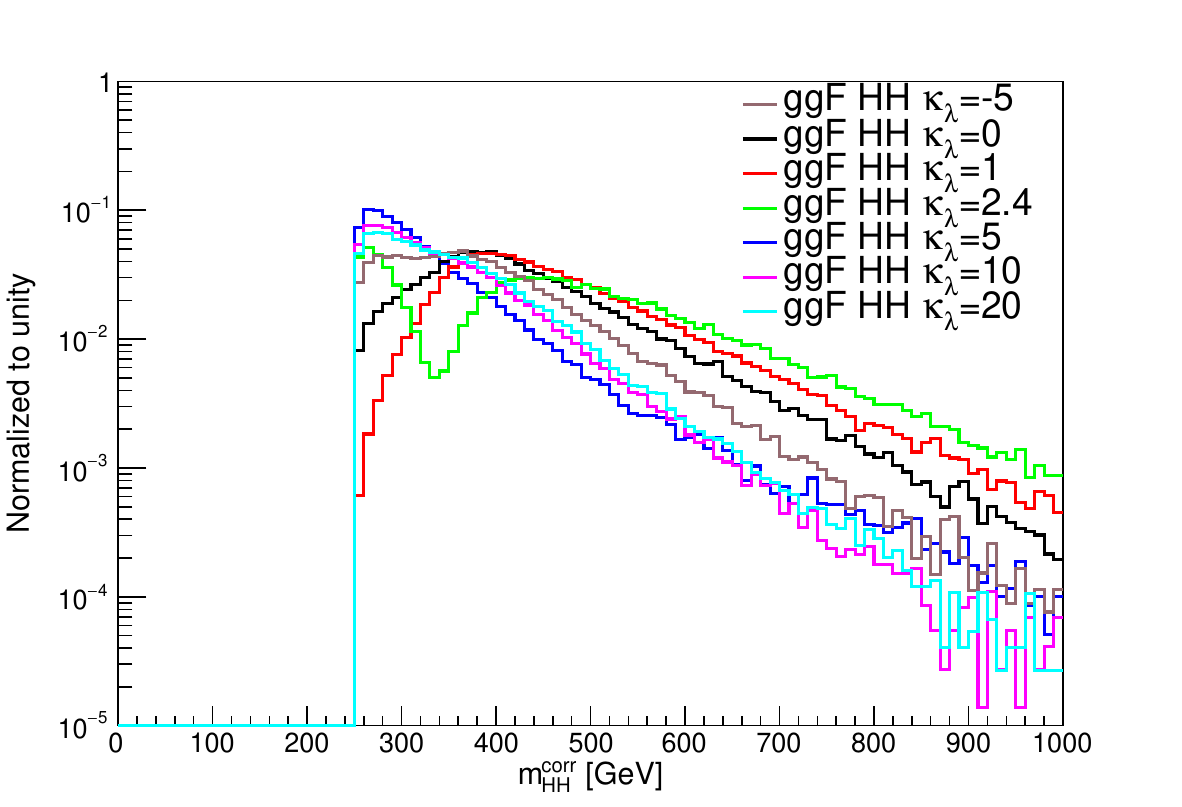}
    \caption{The $m_{\rm{HH}}^{\rm{corr}}$ distribution of ggF signals with different $\kappa_{\lambda}$ values.}
    \label{fig:figure_7}
\end{figure}

\begin{figure}[hbt!]
    \centering
    \includegraphics[width=0.48\textwidth]{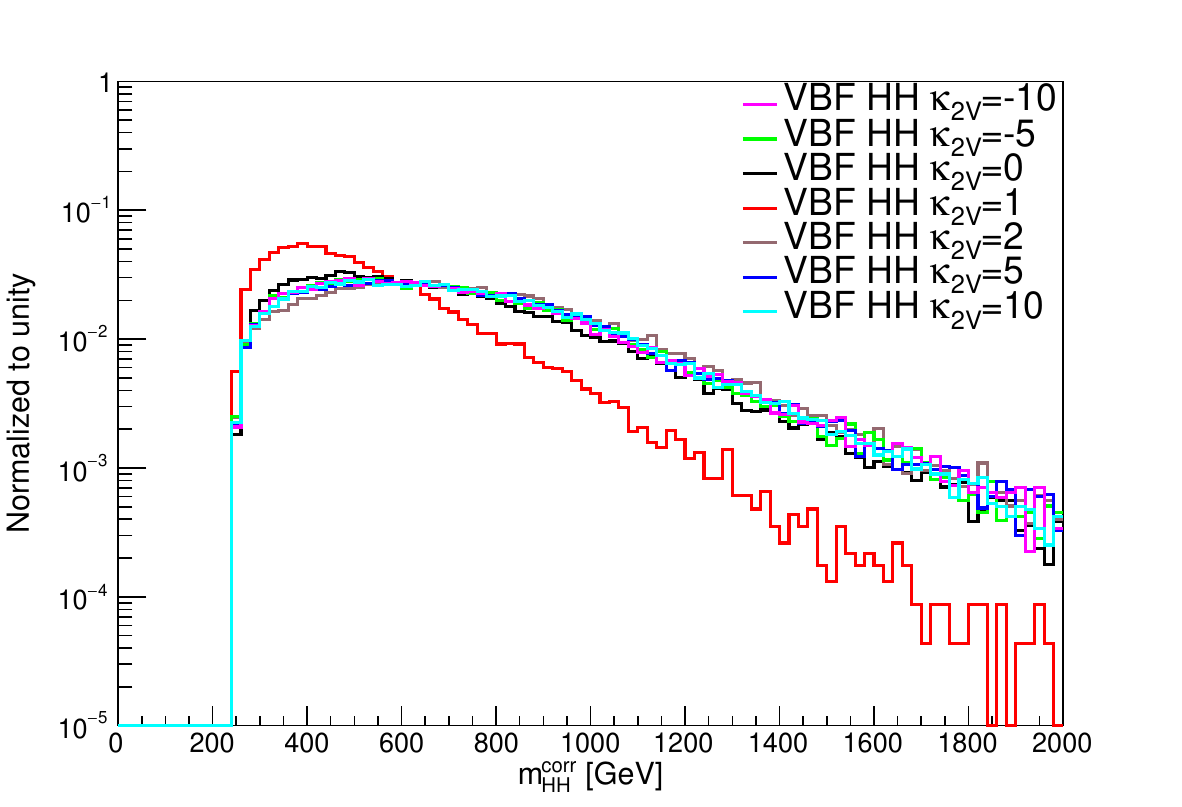}
    \caption{The $m_{\rm{HH}}^{\rm{corr}}$ distribution of VBF signals with different $\kappa_{2\rm{V}}$ values.}
    \label{fig:figure_8}
\end{figure}

In maximizing the sensitivity on the couplings, two categories are defined in each of the ggF and VBF categories. The variable $m_{\rm{HH}}^{\rm{corr}}$ as defined in Eq.~(\ref{eq1})
\begin{equation} \label{eq1}
    m_{\rm{HH}}^{\rm{corr}} = m_{\rm{HH}}-(m_{\mu\mu}-125)-(m_{bb}-125)
\end{equation}
\noindent following Ref.~\cite{CMS:2018tla} is often used in experiments and it serves as a good proxy of $m_{\rm{HH}}$ reflecting the $\kappa_\lambda$ variation, comparing FIG.~\ref{fig:figure_2} and~\ref{fig:figure_7}. It reduces the systematic uncertainties from the energy scale and resolution than directly using the bare $m_{\rm{HH}}$. Maximizing the separation between SM and BSM couplings, a threshold on $m_{\rm{HH}}^{\rm{corr}}$ is determined to divide the ggF events into the ggF SM category with $m_{\rm{HH}}^{\rm{corr}} > $ 400 GeV for a harder $m_{\rm{HH}}^{\rm{corr}}$ spectrum largely contributed by the box diagram, and the ggF BSM category with $m_{\rm{HH}}^{\rm{corr}} < $ 400 GeV for the less energetic events contributed mostly from the triangle diagram that contains the self-coupling $\kappa_\lambda$.
In the VBF category, a further split is optimized for $\kappa_{\text{2V}}$. The $m_{\rm{HH}}^{\rm{corr}}$ varies rapidly when $\kappa_{\text{2V}}$ starts to deviate from its SM value, making a clear difference between the SM and the BSM cases, as shown in FIG.~\ref{fig:figure_8}. Maximizing the separation, the threshold $m_{\rm{HH}}^{\rm{corr}} < $ 680 GeV selects events into the VBF SM category, and the opposite defines the VBF BSM category. The thresholds used in the categorization are summarised in TAB.~\ref{table:2}.


\begin{figure*}[hbt!]
   \centering
   \subcaptionbox{The $|\Delta\eta_{\rm{HH}}|$ distribution in the ggF SM category.\label{fig:figure_9}}
     {\includegraphics[width=0.45\textwidth]{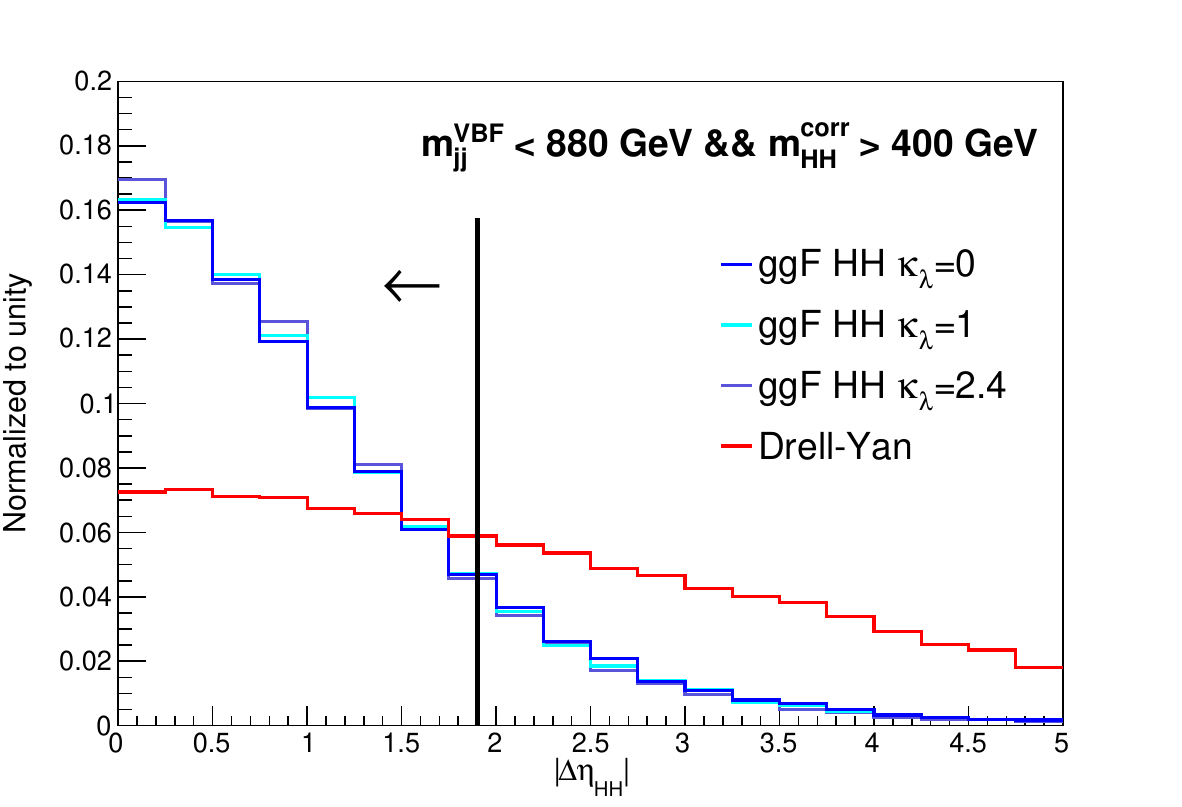}}
   \subcaptionbox{The $p_{\rm{T}}^{bb}/m_{bb}$ distribution in the ggF SM category.\label{fig:figure_10}}
     {\includegraphics[width=0.45\textwidth]{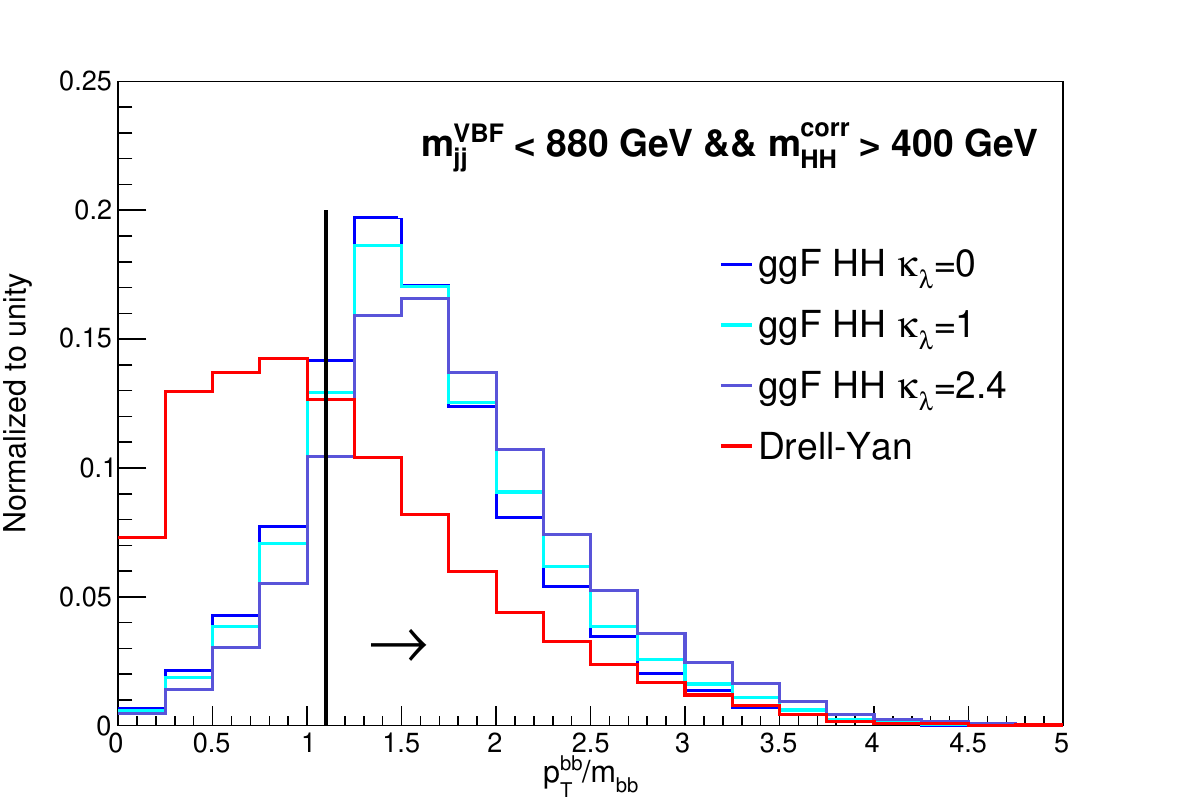}}
   \subcaptionbox{The $p_{\rm{T}}^{\mu\mu} /m_{\mu\mu}$ distribution in the ggF SM category.\label{fig:figure_11}}
     {\includegraphics[width=0.45\textwidth]{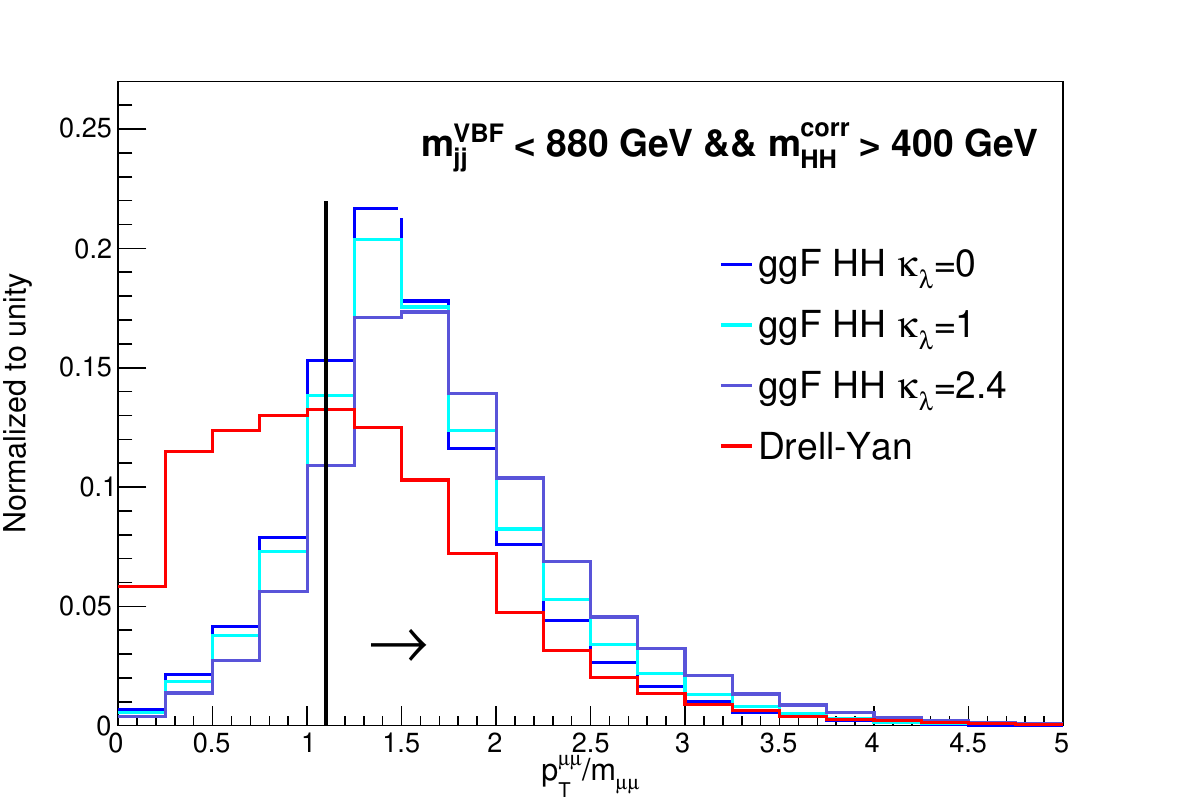}}
   \subcaptionbox{The $p_{\rm{T}}^{bb}/m_{\rm{HH}}$ distribution in the ggF SM category.\label{fig:figure_12}}
     {\includegraphics[width=0.45\textwidth]{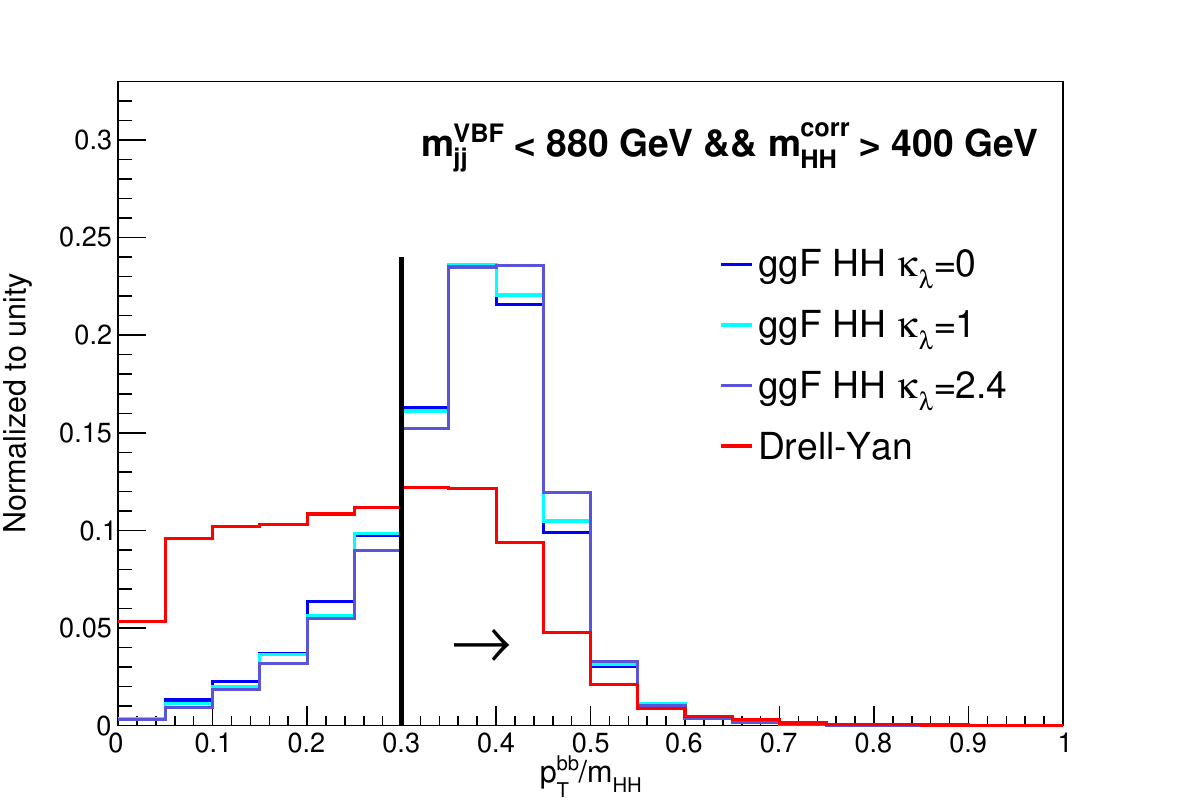}}
   \subcaptionbox{The $H_{\rm{T}}$ distribution in the ggF SM category.\label{fig:figure_13}}
     {\includegraphics[width=0.45\textwidth]{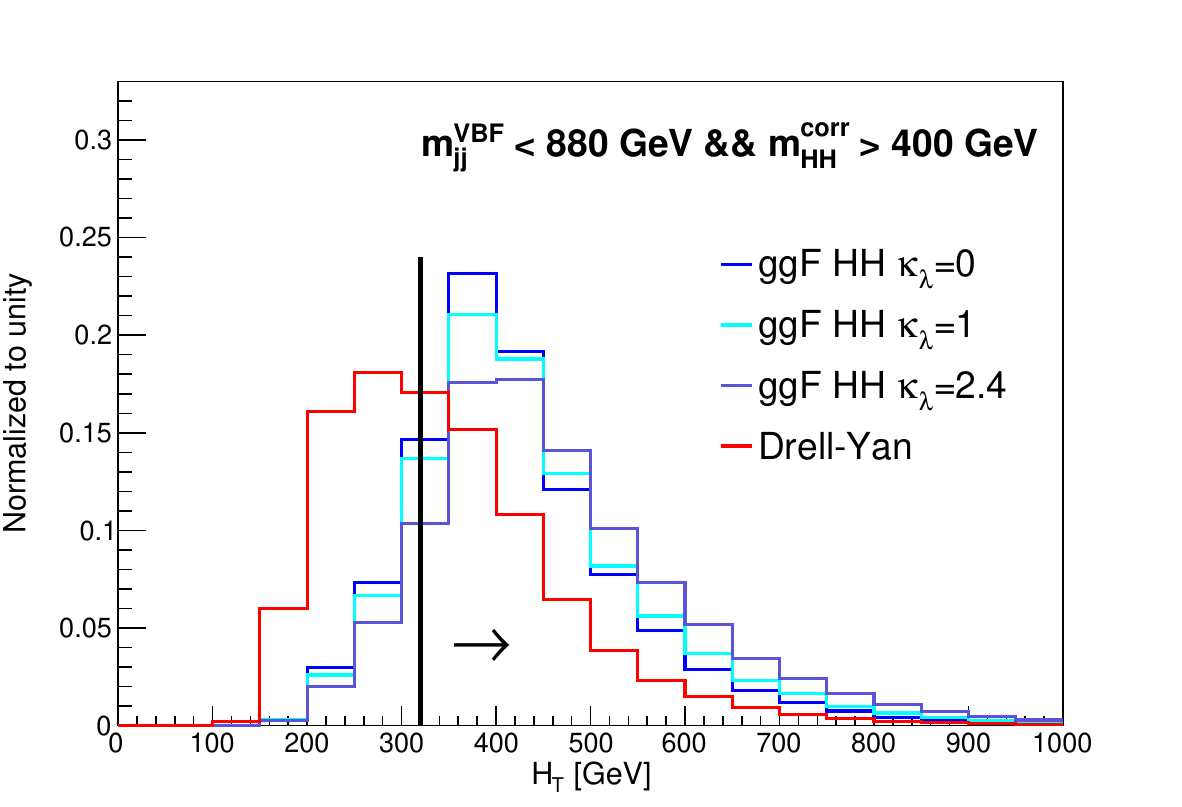}}
   \caption{The discriminating variables chosen in the ggF SM category.}\label{fig:ggF_SM}
\end{figure*}

The categorization based on signal kinematic variations due to the couplings are shared in the cut-based and the BDT analyses. After that, the background amounts and kinematics differ per category. Thus, dedicated cut-based and BDT approaches are optimized or trained separately. In the rest of this section, we focus on the cut-based analysis. The dominant background DY is the main target to suppress. The event selections are optimized by maximizing the expected significance $\rm{S/\sqrt{B}}$ for each category. 

\begin{figure*}[hbt!]
   \centering
   \subcaptionbox{The $|\Delta\eta_{\mu b}^{max}|$ distribution in the ggF BSM category.\label{fig:figure_14}}
     {\includegraphics[width=0.45\textwidth]{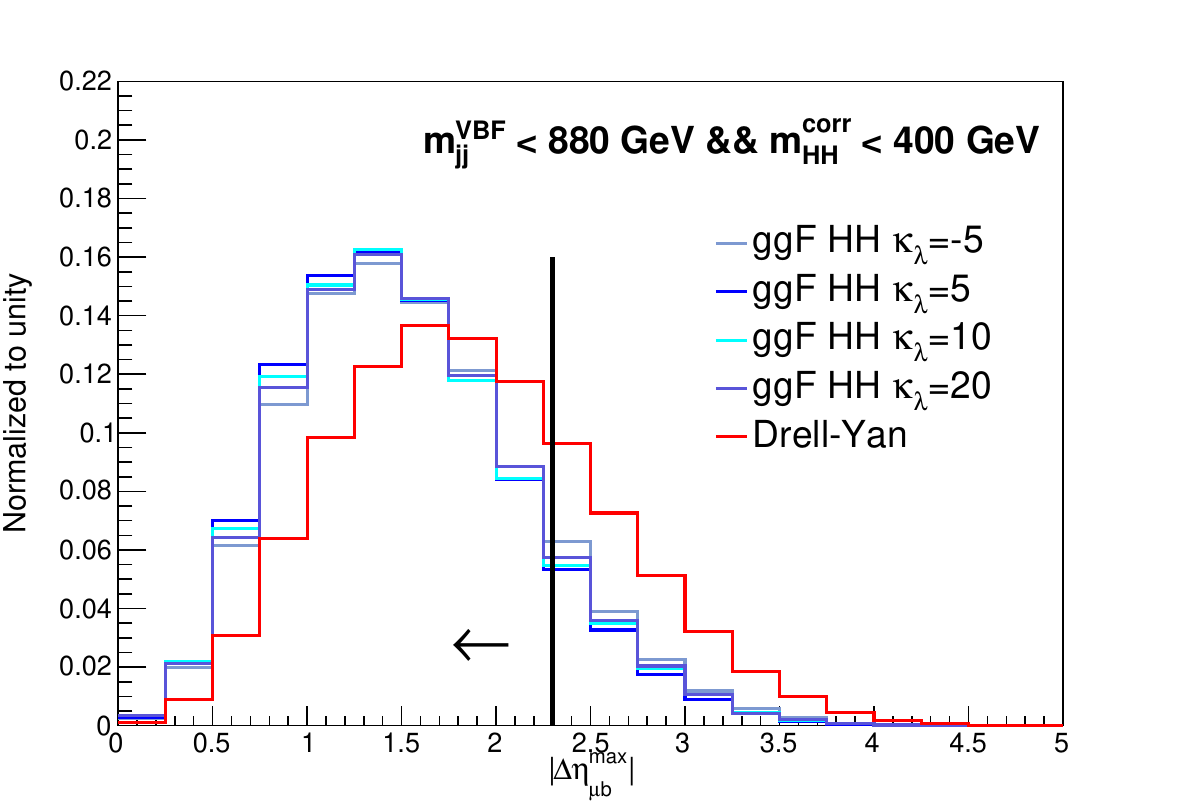}}
   \subcaptionbox{The $|\Delta\eta_{bb}|$ distribution in the ggF BSM category.\label{fig:figure_15}}
     {\includegraphics[width=0.45\textwidth]{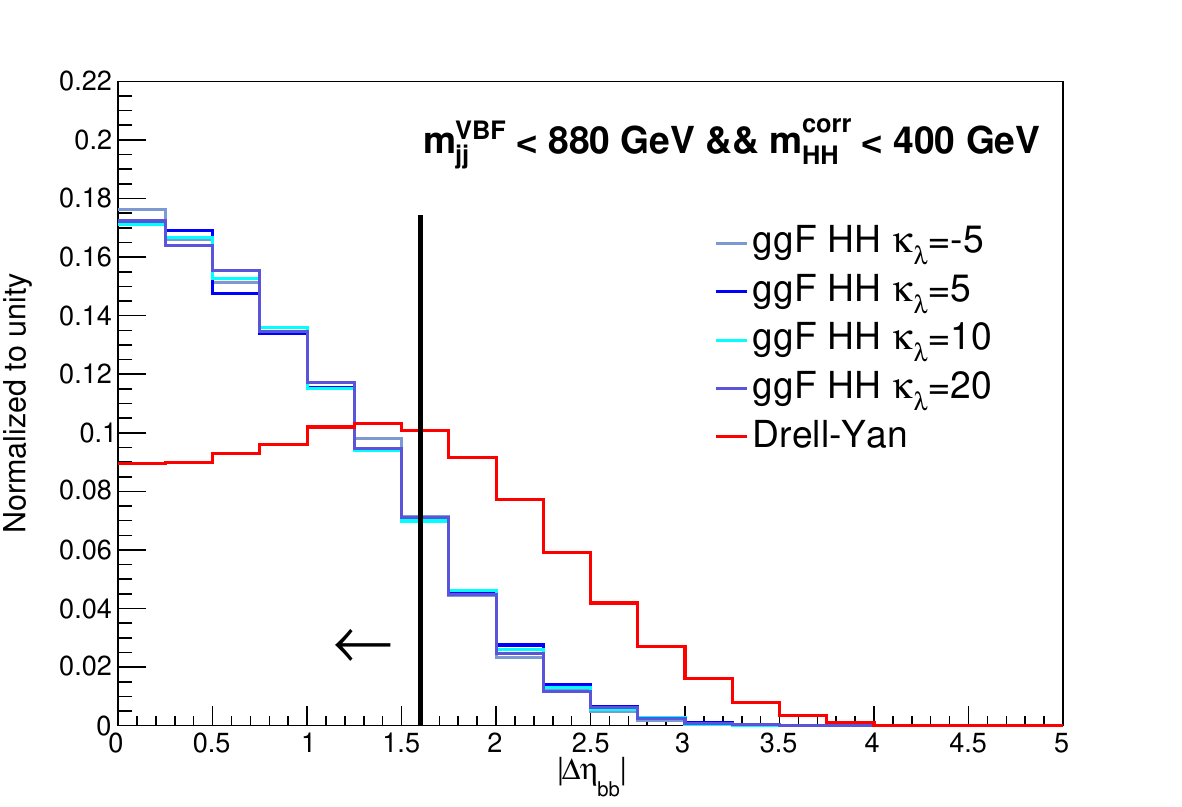}}
   \subcaptionbox{The $E_{\text{T}}^{miss}$ distribution in the ggF BSM category.\label{fig:missET}}
     {\includegraphics[width=0.45\textwidth]{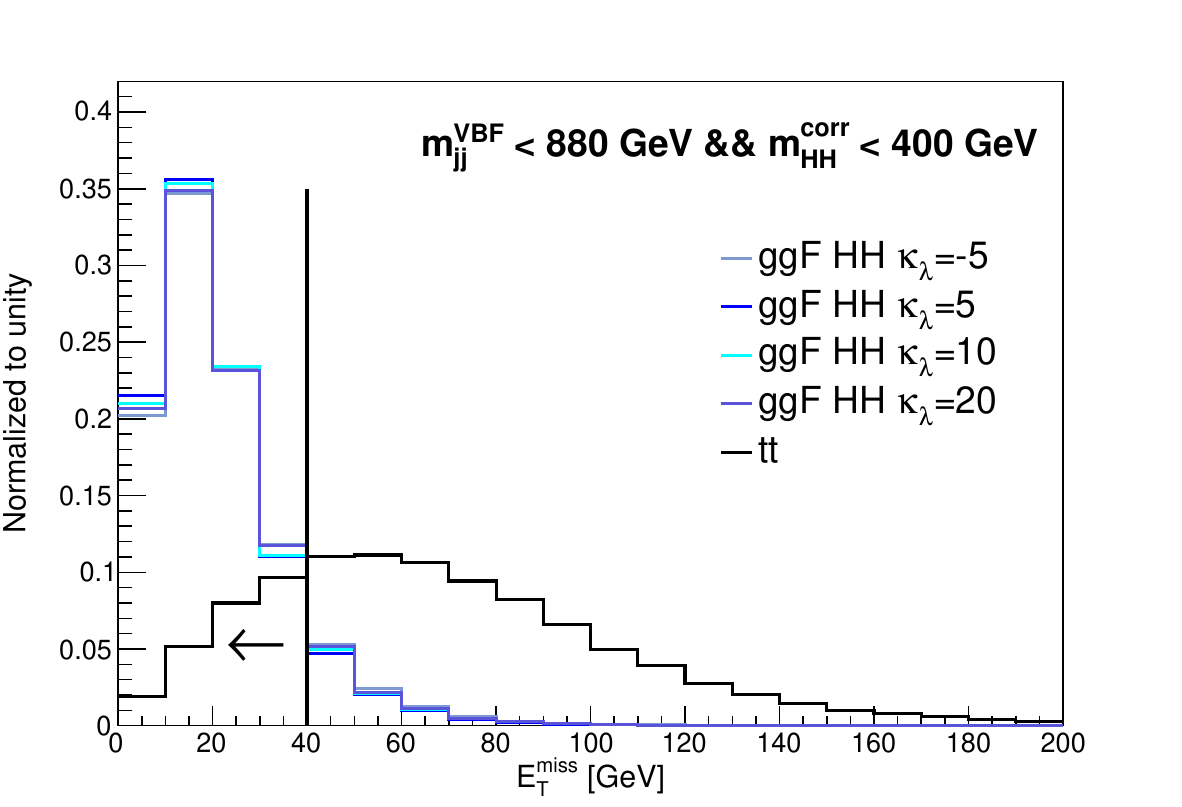}}
   \caption{The discriminating variables chosen in the ggF BSM category.}\label{fig:ggF_BSM}
\end{figure*}

\begin{figure*}[hbt!]
   \centering
   \subcaptionbox{The $m_{\rm{HH}}^{\rm{corr}}$ distribution in the VBF SM category.\label{fig:figure_17}}
     {\includegraphics[width=0.45\textwidth]{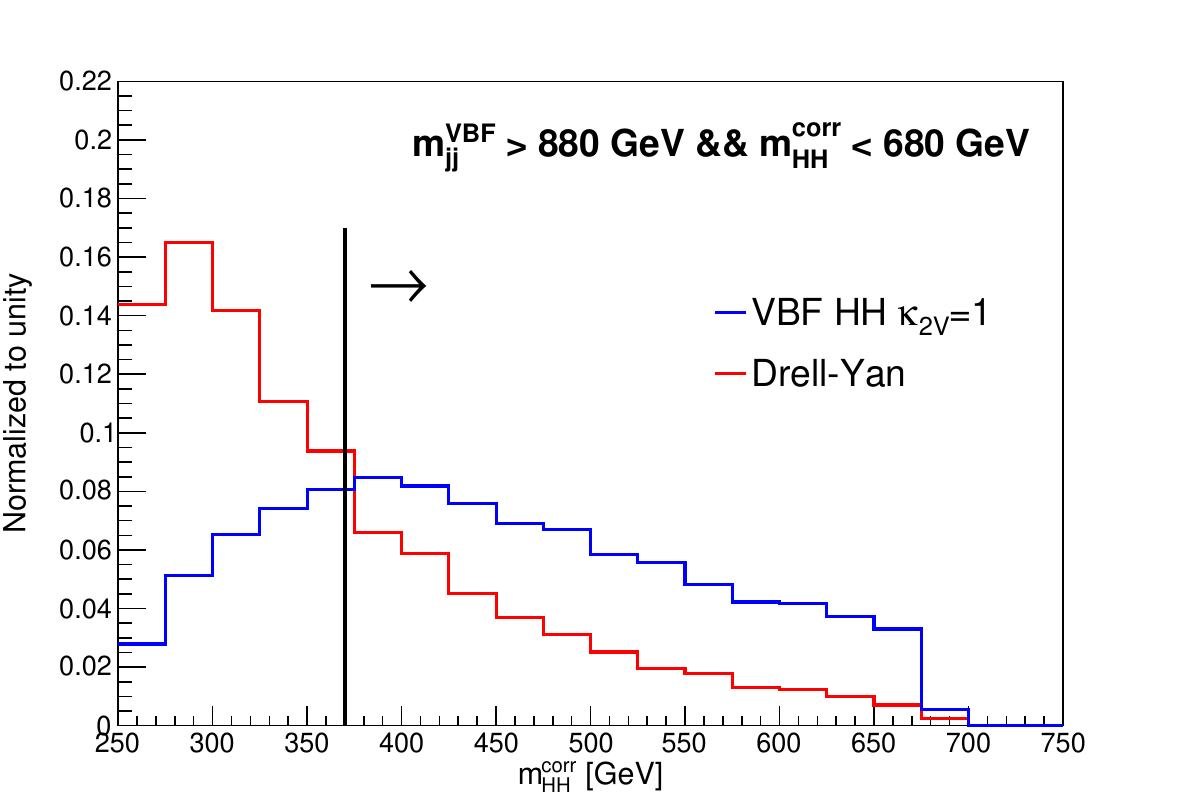}}
   \subcaptionbox{The $|\Delta\eta_{\rm{HH}}|$ distribution in the VBF SM category.\label{fig:figure_18}}
     {\includegraphics[width=0.45\textwidth]{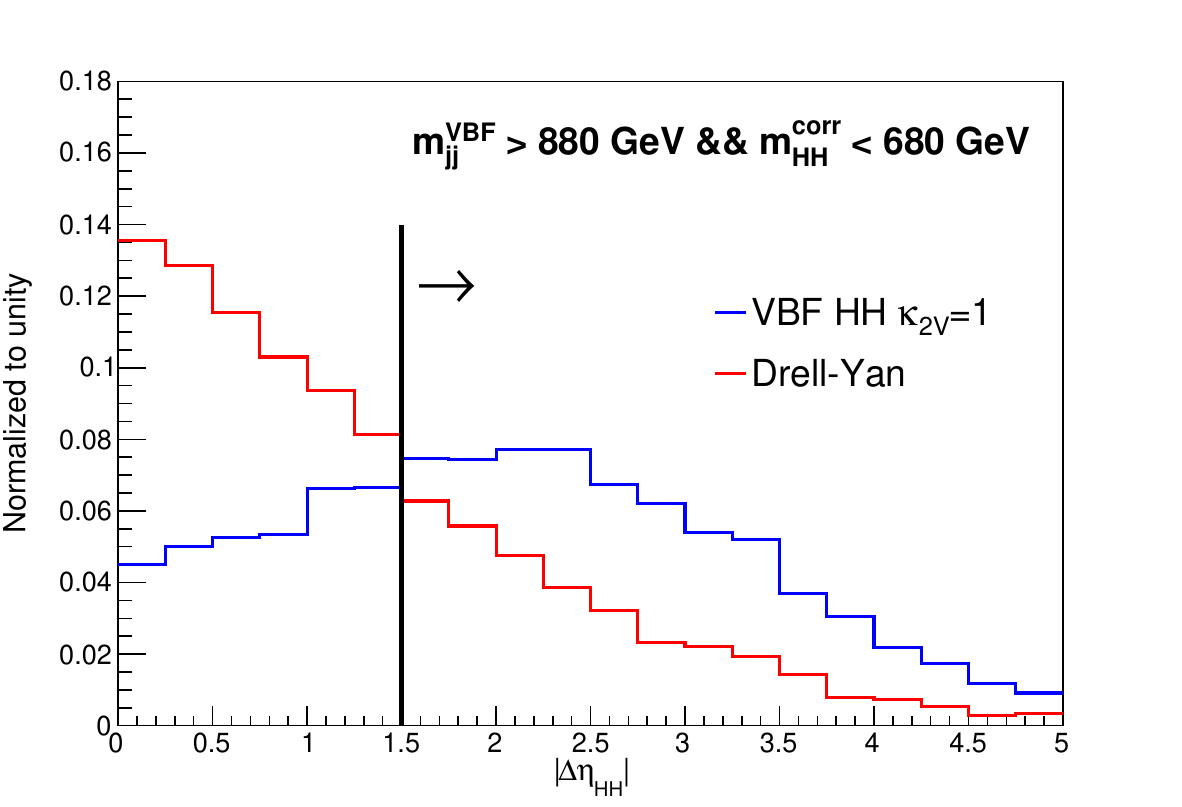}}
   \subcaptionbox{The $|\Delta\eta_{bb}|$ distribution in the VBF SM category.\label{fig:figure_19}}
     {\includegraphics[width=0.45\textwidth]{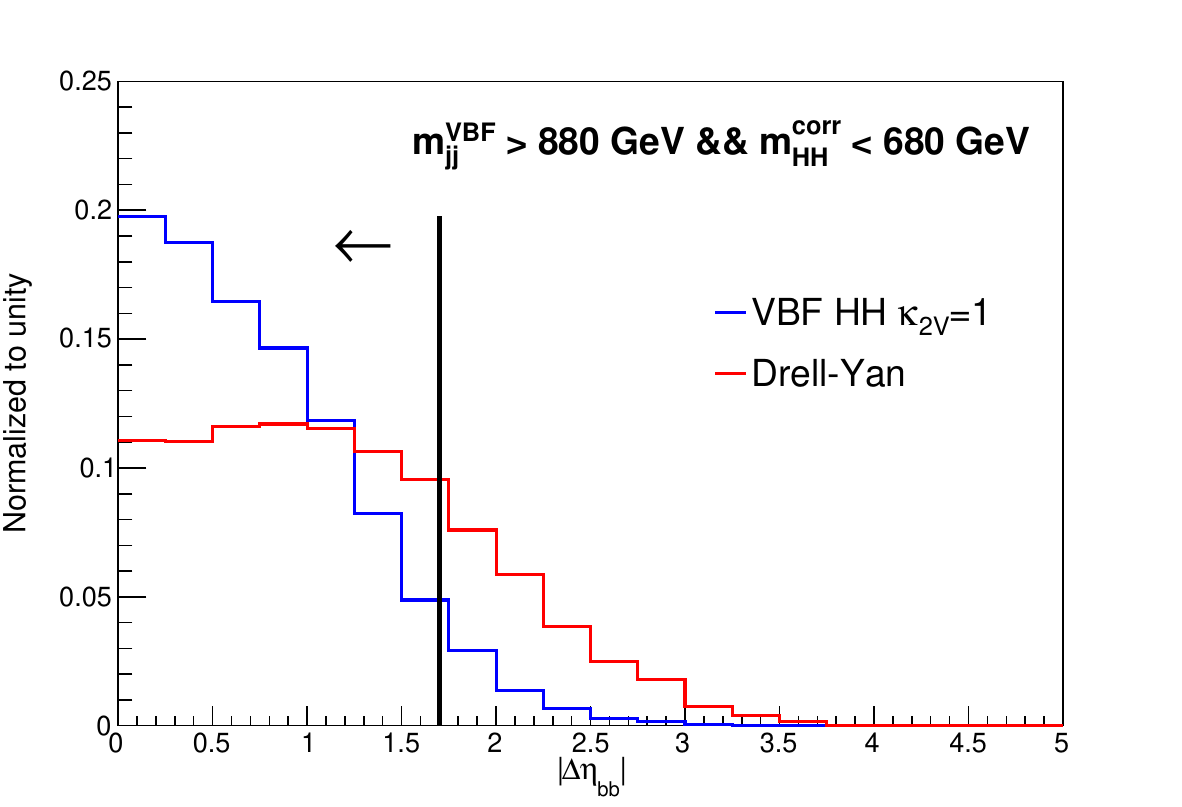}}
   \subcaptionbox{The $p_{\rm{T}}^{bb}/m_{\rm{HH}}$ distribution in the ggF SM category.\label{fig:figure_20}}
     {\includegraphics[width=0.45\textwidth]{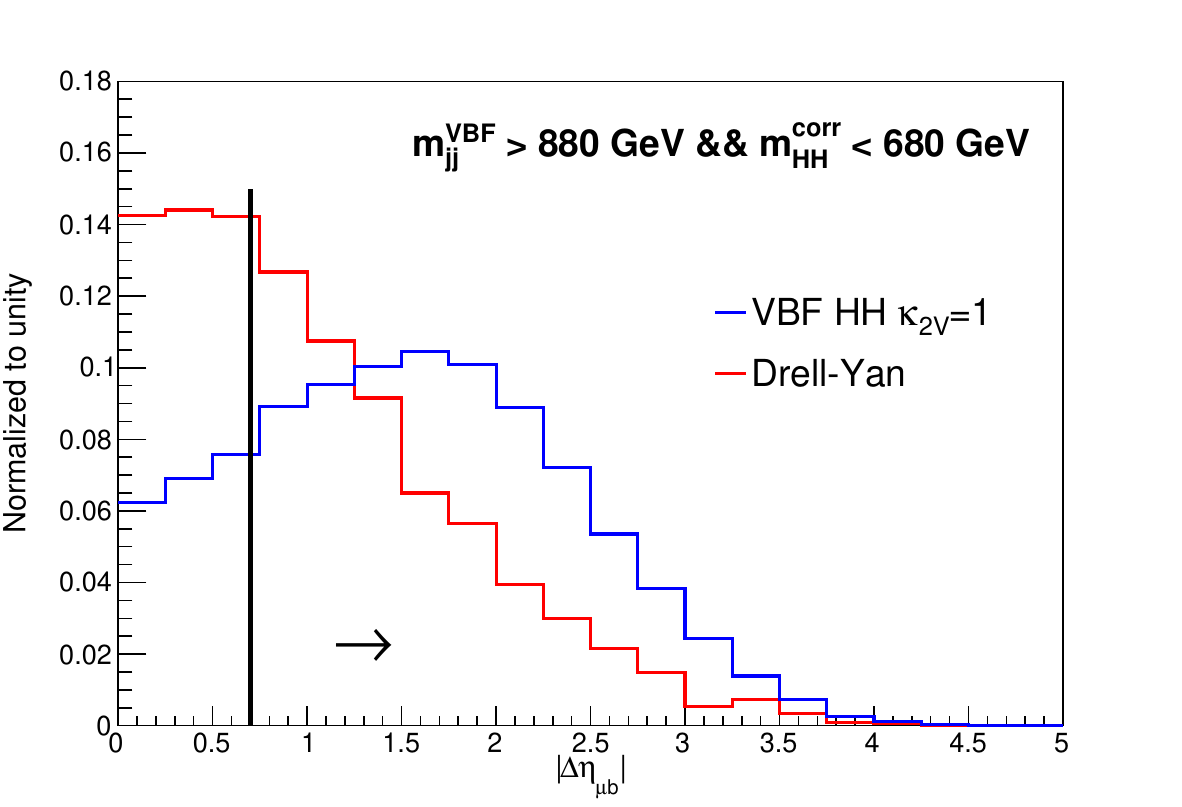}}
   \caption{The discriminating variables chosen in the VBF SM category.}\label{fig:VBF_SM}
\end{figure*}


\begin{figure}[hbt!]
    \centering
    \includegraphics[width=0.48\textwidth]{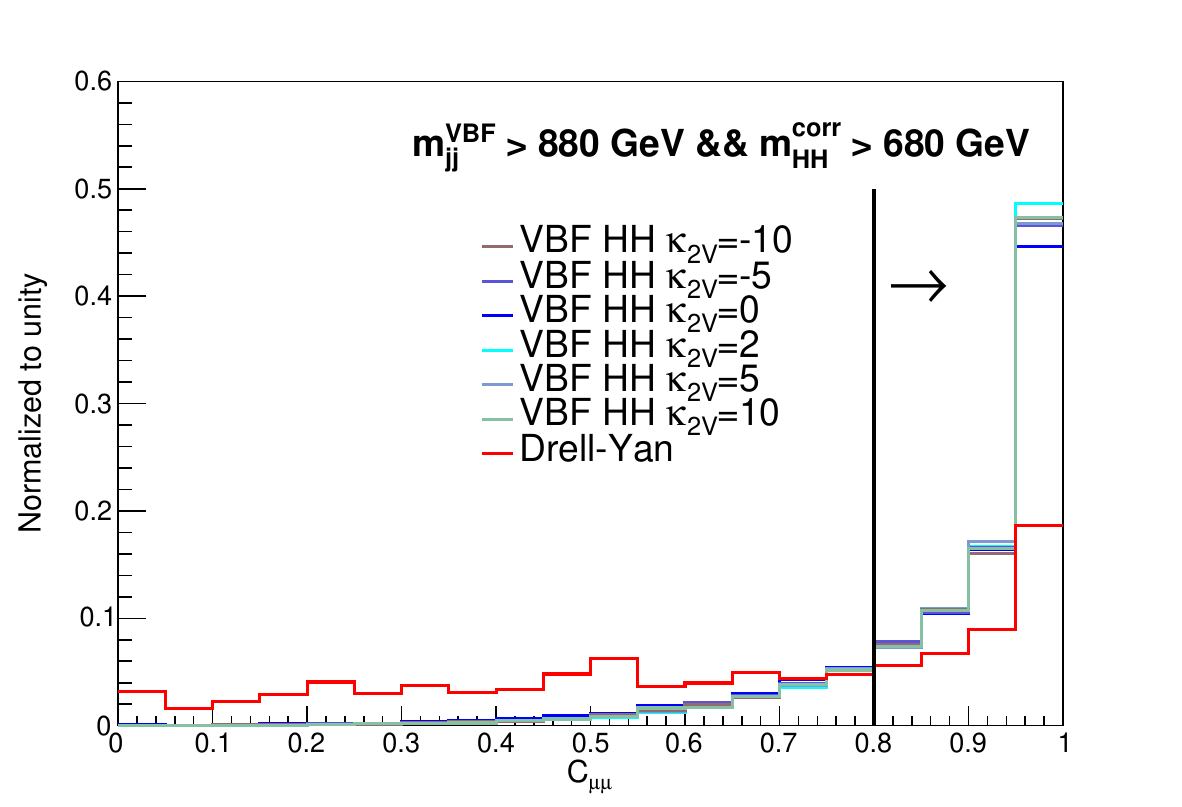}
    \caption{The $C_{(H\to \mu\mu)}$ distribution in the VBF BSM category.}
    \label{fig:figure_16}
\end{figure}

In the ggF SM category, the following discriminating variables are chosen. The $|\Delta\eta_{\rm{HH}}|$ variable stands for the absolute value of the $\eta$ separation between two Higgs bosons. The signal events tend to be more transverse resulting in smaller $|\Delta\eta_{\rm{HH}}|$, while the DY background events are less transverse leading to larger $|\Delta\eta_{\rm{HH}}|$, as shown in FIG.~\ref{fig:figure_9}. The relative $p_{\text{T}}$ variables, such as $p_{\rm{T}}^{bb}/m_{bb}$, $p_{\rm{T}}^{\mu\mu} /m_{\mu\mu}$, $p_{\rm{T}}^{bb}/m_{\rm{HH}}$, where $p_{\rm{T}}^{bb}$ and $p_{\rm{T}}^{\mu\mu}$ are the transverse momenta of the di-bjet and di-muon candidates, can also effectively separate the signal and the DY background events. As shown in FIG.~\ref{fig:figure_10} to~\ref{fig:figure_12}, the signal events tend to have harder spectrum. Lastly, $H_{\rm{T}}$, which represents the scalar sum of the transverse momentum of bjets and muons, has a strong separation power given more energetic signal events, as shown in FIG.~\ref{fig:figure_13}. 

\begin{table*}[htbp!]
\renewcommand{\arraystretch}{1.5}
\centering
\begin{tabular}{ c c c c } 
\hline
\hline
Category\hspace{0.2cm} & Variable Cut\hspace{0.3cm} & $\epsilon_{\text{Signal}}$ & $\epsilon_{\text{DY+tt}}$ \\
\hline
\multirow{5}{5em}{ggF SM}& $|\Delta\eta_{\rm{HH}}|<1.9$ & \multirow{5}{*}{53\% ($\kappa_\lambda=1$)} & \multirow{5}{*}{11\%} \\ 
& $p_{\rm{T}}^{bb}/m_{bb}>1.1$ \\
& $p_{\rm{T}}^{\mu\mu} /m_{\mu\mu}>1.1$ \\
& $p_{\rm{T}}^{bb}/m_{\rm{HH}}>0.3$ \\
& $H_{\rm{T}}>320\ \rm{GeV}$ \\
\hline
\multirow{3}{5em}{ggF BSM}& $|\Delta\eta_{\mu b}^{max}|<2.3$ & \multirow{3}{*}{55\% ($\kappa_\lambda=5$)} & \multirow{3}{*}{13\%}\\
& $|\Delta\eta_{bb}|<1.6$ \\
& $E_{\text{T}}^{miss}<40\ \rm{GeV}$ \\
\hline
\multirow{4}{5em}{VBF SM}& $m_{\rm{HH}}^{\rm{corr}}>370\ \rm{GeV}$ & \multirow{4}{*}{39\% ($\kappa_{\text{2V}}=1$)} & \multirow{4}{*}{3\%} \\
& $|\Delta\eta_{\rm{HH}}|>1.5$ \\
& $|\Delta\eta_{bb}|<1.7$ \\
& $|\Delta\eta_{\mu b}|>0.7$ \\
\hline
VBF BSM & $C_{\mu\mu}>0.8$ & 69\% ($\kappa_{\text{2V}}=10$) & 15\% \\
\hline
\hline
\end{tabular}
\caption{Summary of the optimized cuts for background suppression and the corresponding efficiencies $\epsilon$ in all the four categories. The efficiencies are calculated with the number of events passing the full set of cuts over the number of events that enter the category. For the signal efficiencies, only the relevant signals in the corresponding categories are listed.}
\label{table:cut}
\end{table*}

\begin{table*}[hbt!]
\renewcommand{\arraystretch}{1.5}
\centering
\begin{tabular}{ l c c c c } 
\hline
\hline
\multirow{2}{*}{Process} & \multicolumn{4}{c}{Category} \\
 & ggF SM\hspace{0.15cm} & ggF BSM\hspace{0.15cm} & VBF BSM\hspace{0.15cm} & VBF SM \\
\hline 
ggF HH signal & & & & \\
\hline
$\kappa_{\lambda}=1$ & 1.39 & 0.65 & 0.01 & 0.02 \\
$\kappa_{\lambda}=5$ & 0.71 & 4.27 & 0.004 & 0.01 \\
\hline
VBF HH signal & & & & \\
\hline
$\kappa_{2\rm{V}}=1$ & 0.01 & 0.02 & 0.004 & 0.02 \\
$\kappa_{2\rm{V}}=10$ & 52.5 & 9.24 & 27.9 & 5.61 \\
\hline
Background & & & & \\
\hline
Drell-Yan & 7976 & 54369 & 94.3 & 713 \\
$t\bar{t}$ & 832 & 24619 & 0 & 175 \\
ggH & 1.25 & 1.32 & 0.07 & 0.07\\
VBFH & 0.08 & 0.21 & 0 & 0.04 \\
ZH & 0.74 & 1.32 & 0 & 0.01 \\
ttH & 1.32 & 2.93 & 0.03 & 0.08 \\
bbH & 0.02 & 0.18 & 0 & 0 \\
\hline
Total background & 8811.41 & 78993.96 & 94.40 & 888.20 \\
\hline
\hline
\end{tabular}
\caption{The event yields of signal and background processes in the four categories after background suppression, assuming an integrated luminosity of 3000 fb$^{-1}$.}
\label{table:3}
\end{table*}

In the ggF BSM category, less energetic events are enriched, resulting in very similar behaviour in signal and background events. In this challenging phase space, we find two outstanding variables that can improve the signal significance by calculating $\rm{S/\sqrt{B}}$: $|\Delta\eta_{\mu b}^{max}|$, which is the absolute value of the maximal $\eta$ separation between the muons and the bjets, and $|\Delta\eta_{bb}|$, which stands for the absolute value of the $\eta$ separation between two bjets. As shown in FIG.~\ref{fig:figure_14}, the signal events are more centrally produced leading to slightly smaller $|\Delta\eta_{\mu b}^{max}|$ than the background. In FIG.~\ref{fig:figure_15}, the signal events have two bjets originating from the Higgs boson resulting in smaller $|\Delta\eta_{bb}|$ than the background. Different than other categories, there emerge non-negligible $t\bar{t}$ background events. The missing transverse momentum, $E_{\text{T}}^{miss}$ as a proxy of the neutrino $p_{\text{T}}$,  effectively rejects the $t\bar{t}$ events, as shown in FIG.~\ref{fig:missET}.

In the VBF SM category, the statistics is in general low. We find four variables that improve the significance in a sizeable way by calculating $\rm{S/\sqrt{B}}$: $m_{\rm{HH}}^{\rm{corr}}$, $|\Delta\eta_{\rm{HH}}|$, $|\Delta\eta_{bb}|$ and $|\Delta\eta_{\mu b}|$. Their distributions are shown in FIG.~\ref{fig:figure_17} to ~\ref{fig:figure_20}. The VBF BSM category has the lowest statistics among all due to $m_{\rm{HH}}^{\rm{corr}} >$ 680 GeV leaving little room for optimization. Only one variable, the centrality variable $C_{\mu\mu}$, defined in Eq.~(\ref{eq2}) following Ref.~\cite{CMS:2020tkr}, is chosen, as shown in FIG.~\ref{fig:figure_16}. The signal events tend to have larger $\eta$ separation of the two VBF jets resulting in $C_{\mu\mu}$ more close to 1 than the background events.


\begin{equation}\label{eq2}
    C_{\mu\mu}=e^{[-\frac{4}{(\eta_{1}^{\rm{VBF}}-\eta_{2}^{\rm{VBF}})^2}(\eta^{\mu\mu}-\frac{\eta_{1}^{\rm{VBF}}+\eta_{2}^{\rm{VBF}}}{2} )^2]}
\end{equation}
where $\eta_{1}^{\rm{VBF}}$ and $\eta_{2}^{\rm{VBF}}$ are the pseudorapidities of the two VBF jets.


TAB.~\ref{table:cut} summarizes the optimized cuts for background suppression and the corresponding efficiencies in all the four categories, while TAB.~\ref{table:3} lists the event yields of the signal and background processes, assuming an integrated luminosity of 3000 fb$^{-1}$. We also consider other HH processes, $\text{HH}\to b\bar{b} \rm{WW}$ and $b\bar{b}\tau^{+}\tau^{-}$, where both can lead to $b\bar{b}\mu^{+}\mu^{-}$ + $E_{\text{T}}^{miss}$ final states. However, the aforementioned processes’ contributions to the background are negligible due to relatively smaller di-muon invariant mass than that for the signal.

In the boosted regime where the two b-quarks cannot be resolved in two separate small-R jets, merged jets are reconstructed using $R=0.8$ instead. In this case, the final state includes a large-R jet and two muons. However, this region has too low statistics and ends up with negligible contribution with respect to the analysis using resolved b-quarks.

\section{BDT Analysis}
\label{sec:bdt}

In the ggF SM and BSM categories, the sufficient statistics allows to apply machine learning algorithms for further improvement in the sensitivity. The BDT algorithm is adopted using the package of XGBoost~\cite{Chen:2016btl}.

The general strategy is to train a huge amount of shallow decision trees and extract a strong separation power from the tree ensemble. The training setup includes 2500 trees, the tree depth of 3 and a learning rate of 0.08 (0.1) for ggF SM (BSM) category. The shallow trees with only a depth of 3 are proven to be an effective way of avoiding over-training. The MC samples are split into 64\%, 16\% and 20\% for training, testing and application, correspond to the numbers of events of 101k (335k), 25k (84k) and 32k (104k) for the ggF SM (BSM) category. The ratio of signal, DY and $t\bar{t}$ events in all samples is 40:60:1 (4:7:1) for the ggF SM (BSM) category. The proportion of $t\bar{t}$ in the training is low in the ggF SM category given its small contribution. The input variables in the training are listed in TAB.~\ref{table:input}. The training is performed separately in the ggF SM and BSM categories. The signal sample generated with $\kappa_{\lambda}=1$ is used in the training for the ggF SM category, while the signal samples generated with $\kappa_{\lambda}=$~5, 10 and 20 are used in the training for the ggF BSM category. Both the DY and $t\bar{t}$ processes are used as the background in the training in the two categories.

\begin{table*}[htbp!]
\renewcommand{\arraystretch}{1.5}
\centering
\begin{tabular}{ c c c } 
\hline
\hline
\multirow{2}{*}{Input Variable} & \multicolumn{2}{c}{Category} \\
 & ggF SM & ggF BSM \\
\hline
$p_{\rm{T}}^{\mu 1}$,$p_{\rm{T}}^{\mu 2}$,$p_{\rm{T}}^{b 1}$,$p_{\rm{T}}^{b 2}$ & \checkmark & \checkmark\\ 
$E_{\mu 1}$,$E_{\mu 2}$,$E_{b 1}$,$E_{b 2}$ & \checkmark & \\
$\eta^{\mu 1}$,$\eta^{\mu 2}$ & \checkmark & \\
$\eta^{b 1}$,$\eta^{b 2}$ & \checkmark & \checkmark\\
$\eta^{\rm{VBF}}_{j1}$ &  & \checkmark\\
$E_{\mu\mu}$,$E_{bb}$,$\eta_{\mu\mu}$,$\eta_{bb}$,$\cos{\theta_{\mu\mu}}$,$\cos{\theta_{bb}}$ & \checkmark & \\
$p_{\rm{T}}^{\mu\mu}$,$p_{\rm{T}}^{bb}$,$m_{\mu\mu}$,$m_{bb}$ & \checkmark & \checkmark\\
$m_{\rm{HH}}$,$m_{\rm{HH}}^{\rm{corr}}$ & \checkmark & \\
$p_{\rm{T}}^{b 1} /m_{bb}$,$p_{\rm{T}}^{b 2} /m_{bb}$,$p_{\rm{T}}^{bb} /m_{bb}$,$p_{\rm{T}}^{\mu 1} /m_{\mu\mu}$,$p_{\rm{T}}^{\mu 2} /m_{\mu\mu}$ & \checkmark & \\
$p_{\rm{T}}^{bb} /m_{bb}$,$p_{\rm{T}}^{bb}/m_{\rm{HH}}$,$p_{\rm{T}}^{\mu\mu} /m_{\mu\mu}$,$p_{\rm{T}}^{\mu\mu}/m_{\rm{HH}}$ & \checkmark & \checkmark\\
$H_{\rm{T}}$,$p_{\rm{T}}^{\rm{HH}}$,$p_{\rm{T}}^{\mu\mu}/p_{\rm{T}}^{bb}$ & \checkmark & \checkmark\\
$E_{\text{T}}^{miss}$,$\eta^{miss}$ & \checkmark & \checkmark\\
$|\Delta\eta_{\rm{HH}}|$,$|\Delta\eta_{\mu b}|$,$|\Delta\eta_{\mu b}^{max}|$,$|\Delta\eta_{\mu b}^{other}|$ & \checkmark & \checkmark\\
$|\Delta\eta_{bb}|$,$|\Delta\eta_{\mu\mu}|$ & \checkmark & \checkmark\\
$|\Delta R_{\rm{HH}}|$,$|\Delta R_{\mu b}|$,$|\Delta R_{bb}|$,$|\Delta R_{\mu\mu}|$ & \checkmark & \checkmark\\
$|\Delta R_{\mu b}^{min}|$,$|\Delta R_{\mu b}^{other}|$,$|\Delta R_{jj}^{\rm{VBF}}|$ & \checkmark & \\
$|\Delta\phi_{\rm{HH}}|$,$|\Delta\phi_{\mu b}|$,$|\Delta\phi_{bb}|$,$|\Delta \phi_{jj}^{\rm{VBF}}|$ & \checkmark & \checkmark\\
$|\Delta\phi_{\mu\mu}|$ & \checkmark & \\
\hline
\hline
\end{tabular}
\caption{Summary of input variables for the BDT training in the two ggF categories. Besides the variables that are already explained in the texts, $|\Delta\eta_{\mu b}^{max}|$ is the maximal $|\Delta\eta|$ between muons and bjets, while $|\Delta\eta_{\mu b}^{other}|$ is for the other muon and bjet. $|\Delta R_{\mu b}^{min}|$ and $|\Delta R_{\mu b}^{other}|$ are defined accordingly.}
\label{table:input}
\end{table*}

\begin{figure*}[hbt!]
  \centering
  \subcaptionbox{The ggF SM category\label{fig:sm_output}}
     {\includegraphics[width=0.45\textwidth]{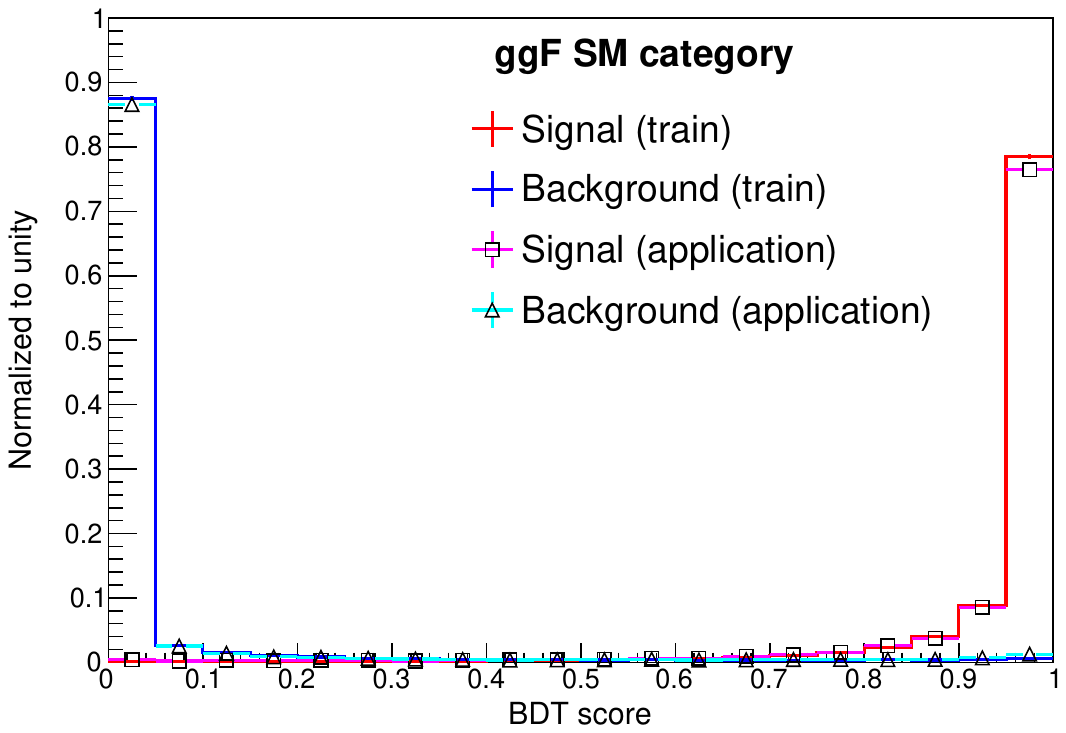}}
  \subcaptionbox{The ggF BSM category\label{fig:bsm_output}}
     {\includegraphics[width=0.45\textwidth]{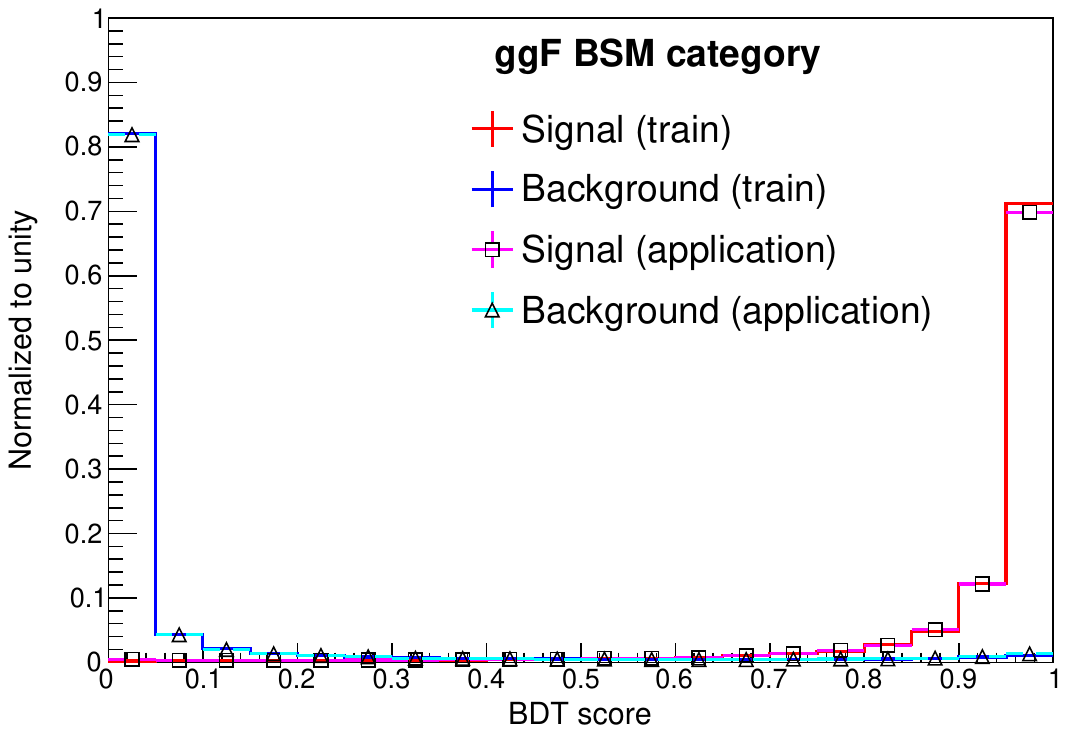}}
  \caption{The BDT score distributions in the ggF SM and BSM categories.}\label{fig:ggF_BDT}
\end{figure*}
\begin{figure*}[hbt!]
  \centering
  \subcaptionbox{The ggF SM category\label{fig:ggfsm_ROC}}
     {\includegraphics[width=0.45\textwidth]{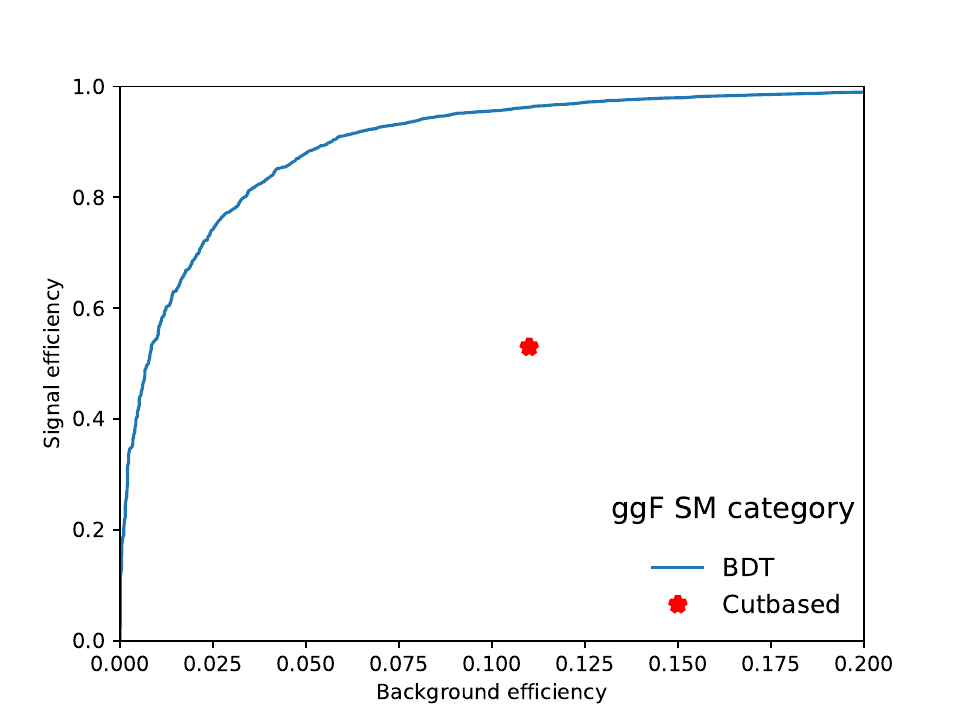}}
  \subcaptionbox{The ggF BSM category\label{fig:ggfbsm_ROC}}
     {\includegraphics[width=0.45\textwidth]{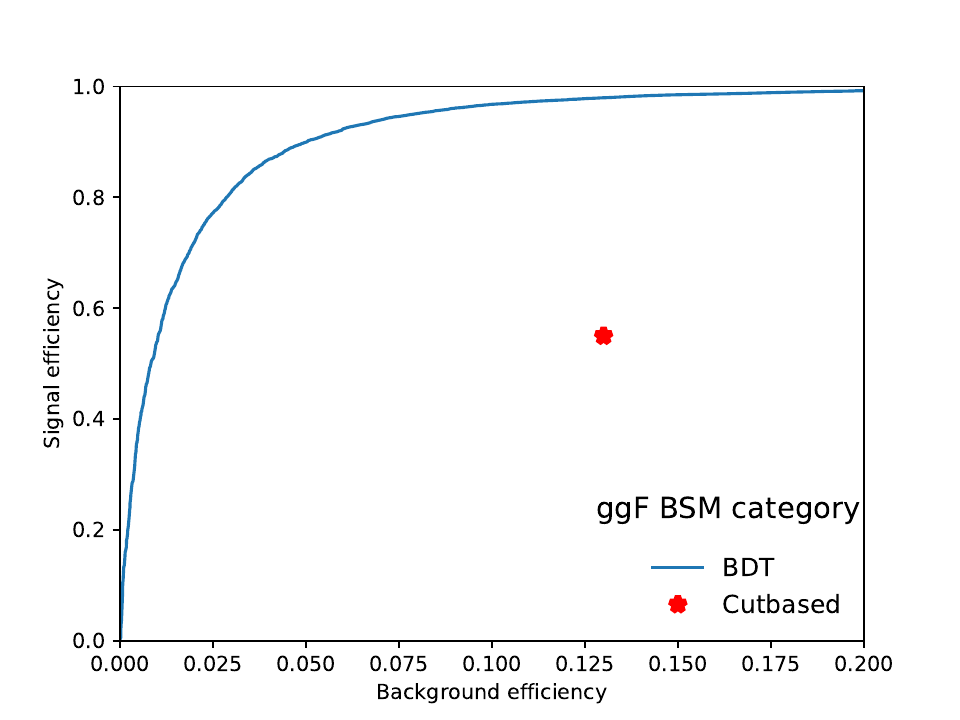}}
  \caption{The ROC curve in the training of the ggF SM and BSM categories. The red dots present the performance of the cut-based approach.}\label{fig:ggF_ROC}
\end{figure*}


The BDT score distributions of the training and testing samples are compared and a good agreement is found, which suggests no over-training issue. The BDT score distributions are then shown with the training and the application samples, the latter of which are used for the inference in the analysis, in FIG.~\ref{fig:sm_output} and FIG.~\ref{fig:bsm_output} for the ggF SM and BSM categories, respectively. Still, there is found no over-training issue.

A cut on the BDT score is applied to purify the signal. The threshold is chosen to keep the signal efficiency roughly equal to the one in the cut-based analysis which is around 50\% as listed in TAB.~\ref{table:cut}. With this threshold, the background efficiency is much suppressed down to 0.85\% and 0.94\% for the ggF SM and BSM categories, respectively, one order of magnitude smaller than the ones from the cut-based analysis as listed in TAB.~\ref{table:cut}. The improvement by BDT is also visualized in the ROC curves in FIG.~\ref{fig:ggfsm_ROC} and~\ref{fig:ggfbsm_ROC}. The cut-based performance is shown as the red stars for comparisons. The events with the BDT score above the threshold are used in the final fits described in the next Section.

\section{Results}
\label{sec:res}

\begin{figure*}[hbt!]
   \centering
   \subcaptionbox{The distribution of $m_{\mu\mu}$ in the ggF SM category.\label{fig:m2mu_ggFSM}}
     {\includegraphics[width=0.45\textwidth]{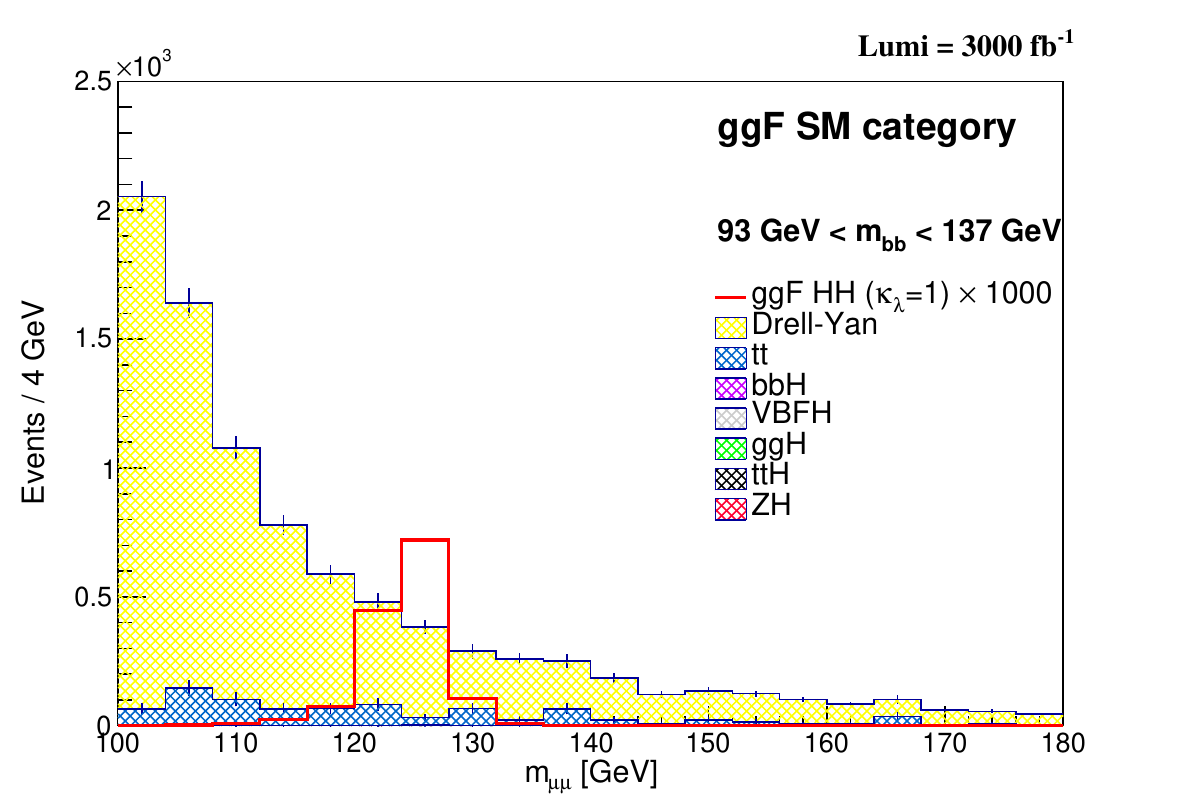}}
   \subcaptionbox{The distribution of $m_{bb}$ in the ggF SM category.\label{fig:m2bj_ggFSM}}
     {\includegraphics[width=0.45\textwidth]{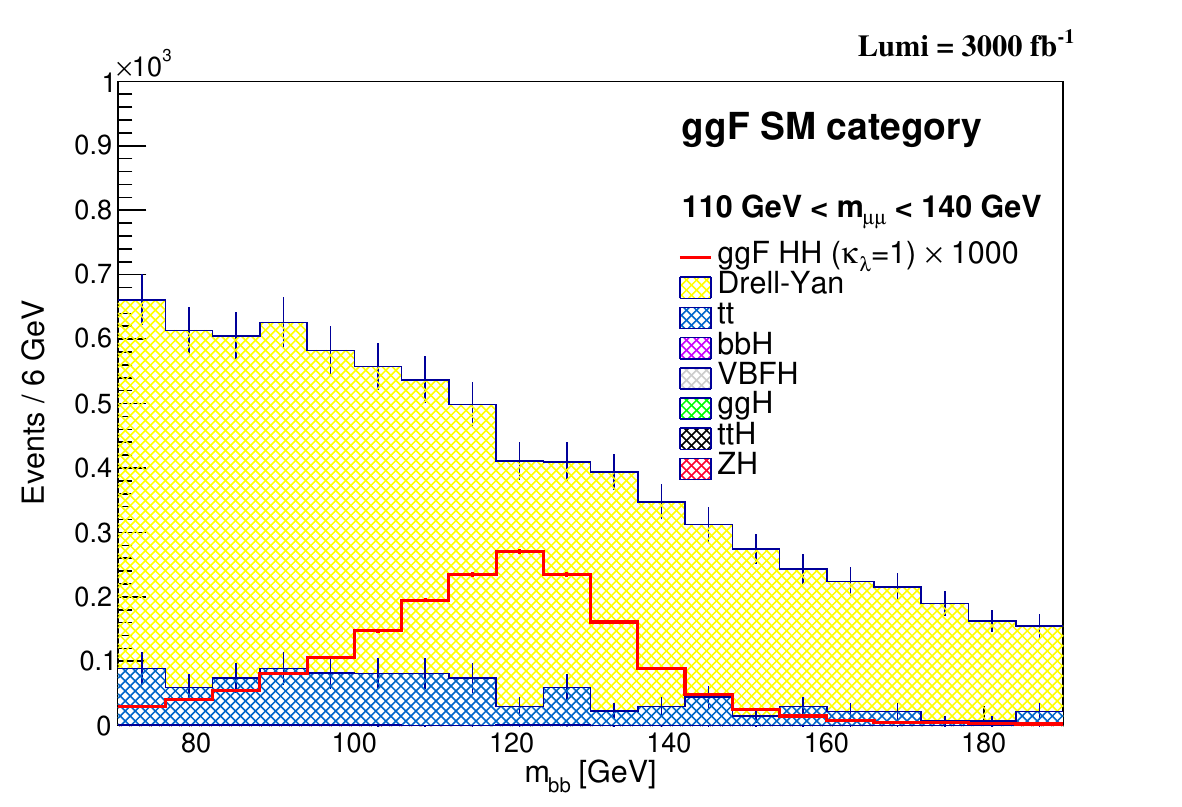}}
   \caption{The distributions of $m_{\mu\mu}$ and $m_{bb}$ in the ggF SM category from the cut-based analysis.}\label{fig:ggFSM_2mu2bj}
\end{figure*}

\begin{figure*}[hbt!]
   \centering
   \subcaptionbox{The distribution of $m_{\mu\mu}$ in the ggF BSM category.\label{fig:m2mu_ggFBSM}}
     {\includegraphics[width=0.45\textwidth]{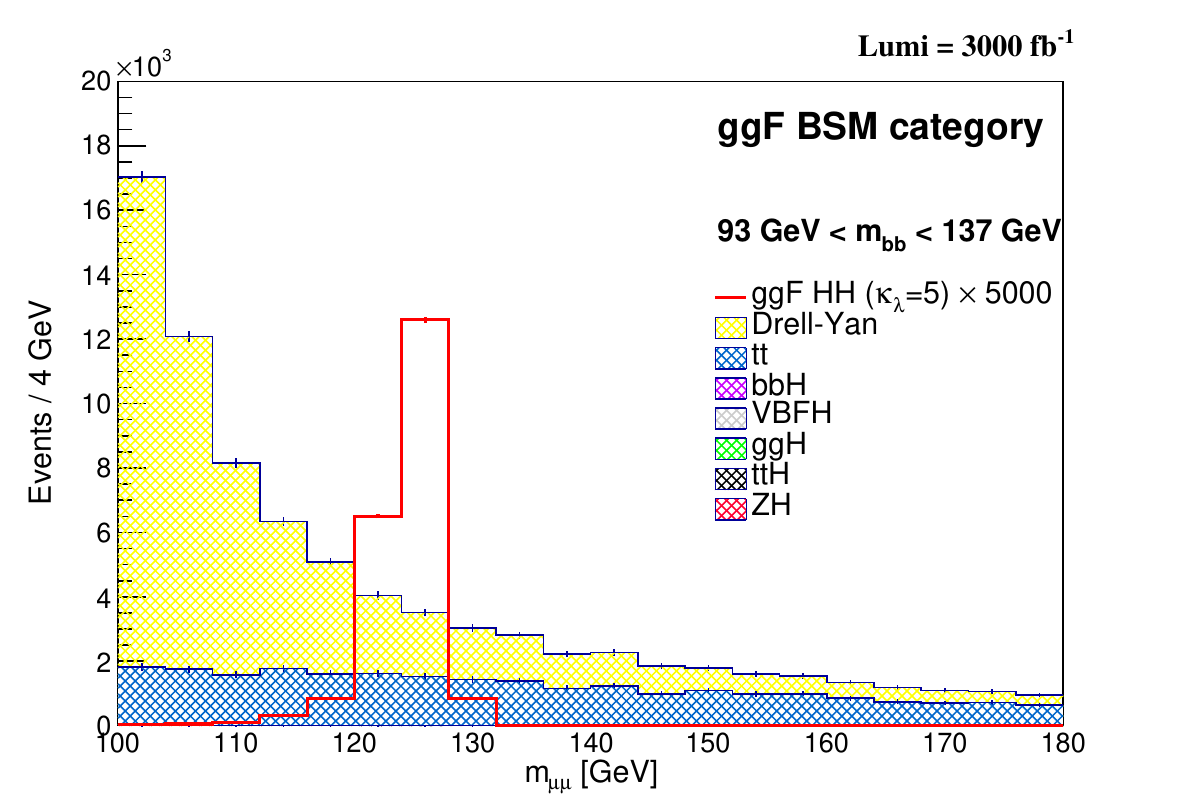}}
   \subcaptionbox{The distribution of $m_{bb}$ in the ggF BSM category.\label{fig:m2bj_ggFBSM}}
     {\includegraphics[width=0.45\textwidth]{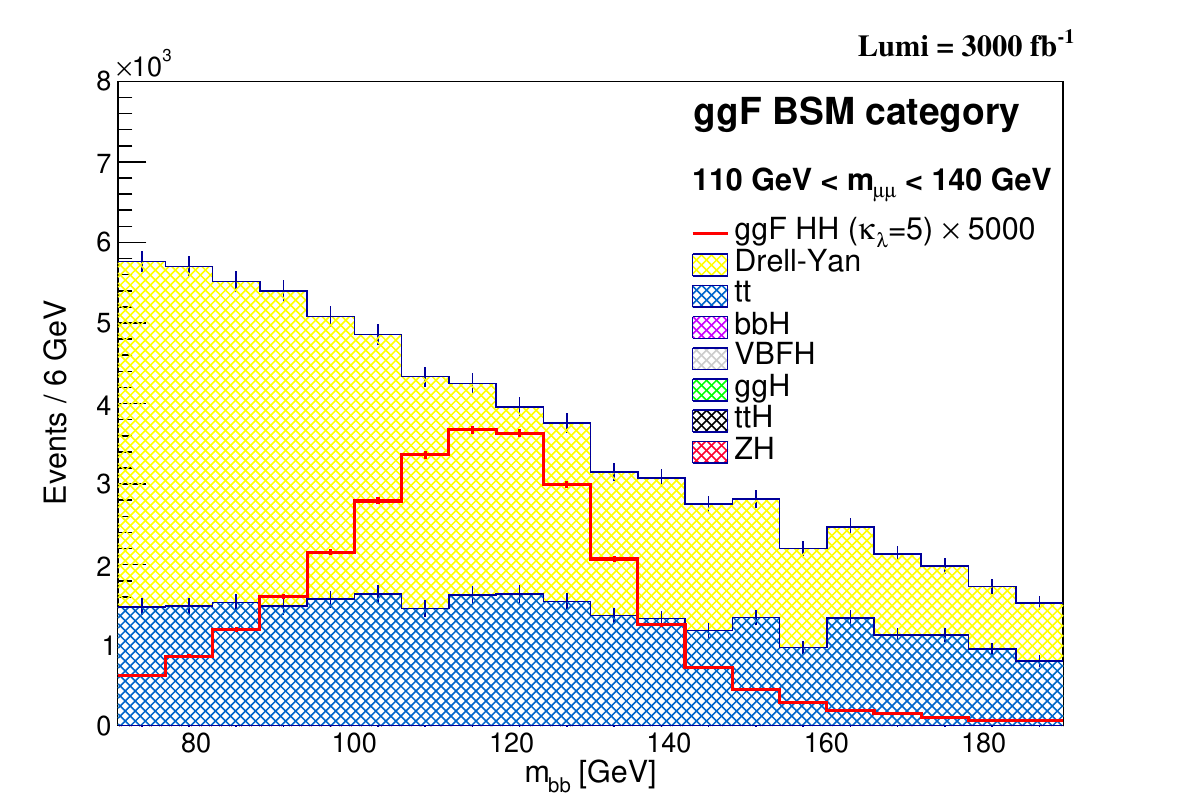}}
   \caption{The distributions of $m_{\mu\mu}$ and $m_{bb}$ in the ggF BSM category from the cut-based analysis.}\label{fig:ggFBSM_2mu2bj}
\end{figure*}

\begin{figure*}[hbt!]
   \centering
   \subcaptionbox{The distribution of $m_{\mu\mu}$ in the VBF BSM category.\label{fig:m2mu_VBFBSM}}
     {\includegraphics[width=0.45\textwidth]{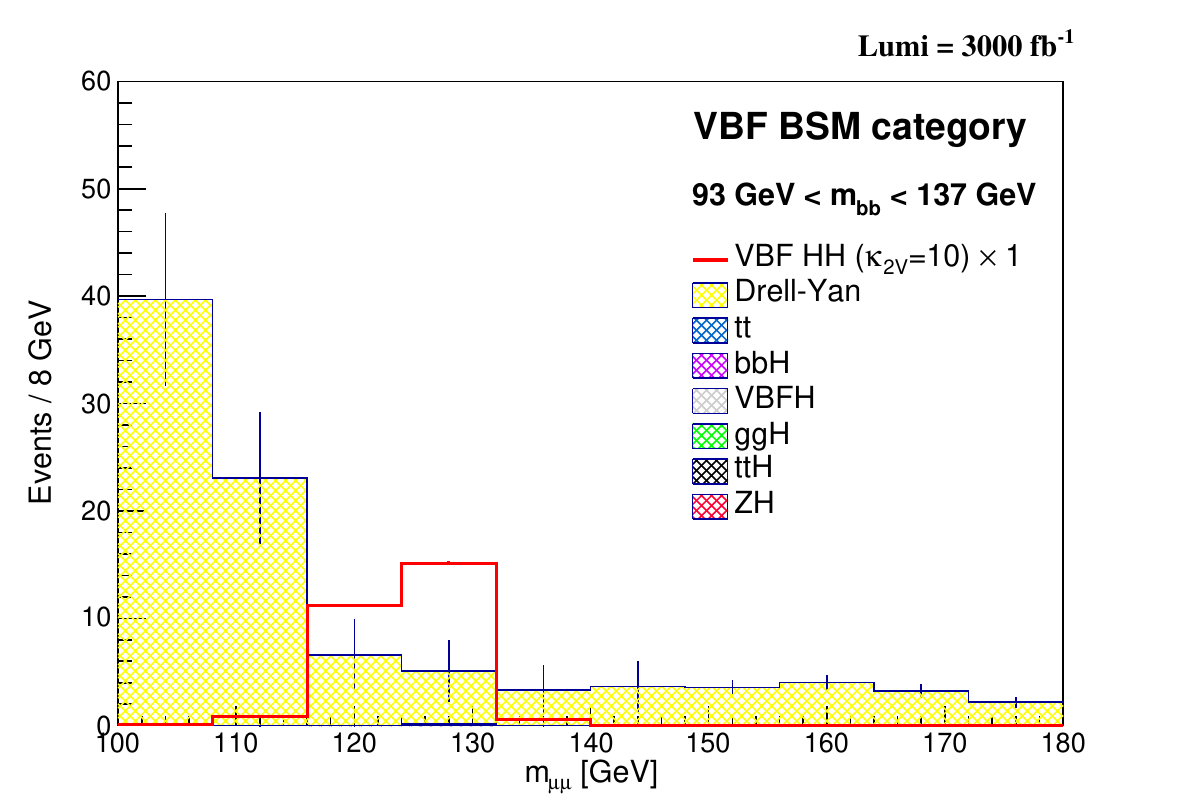}}
   \subcaptionbox{The distribution of $m_{bb}$ in the VBF BSM category.\label{fig:m2bj_VBFBSM}}
     {\includegraphics[width=0.45\textwidth]{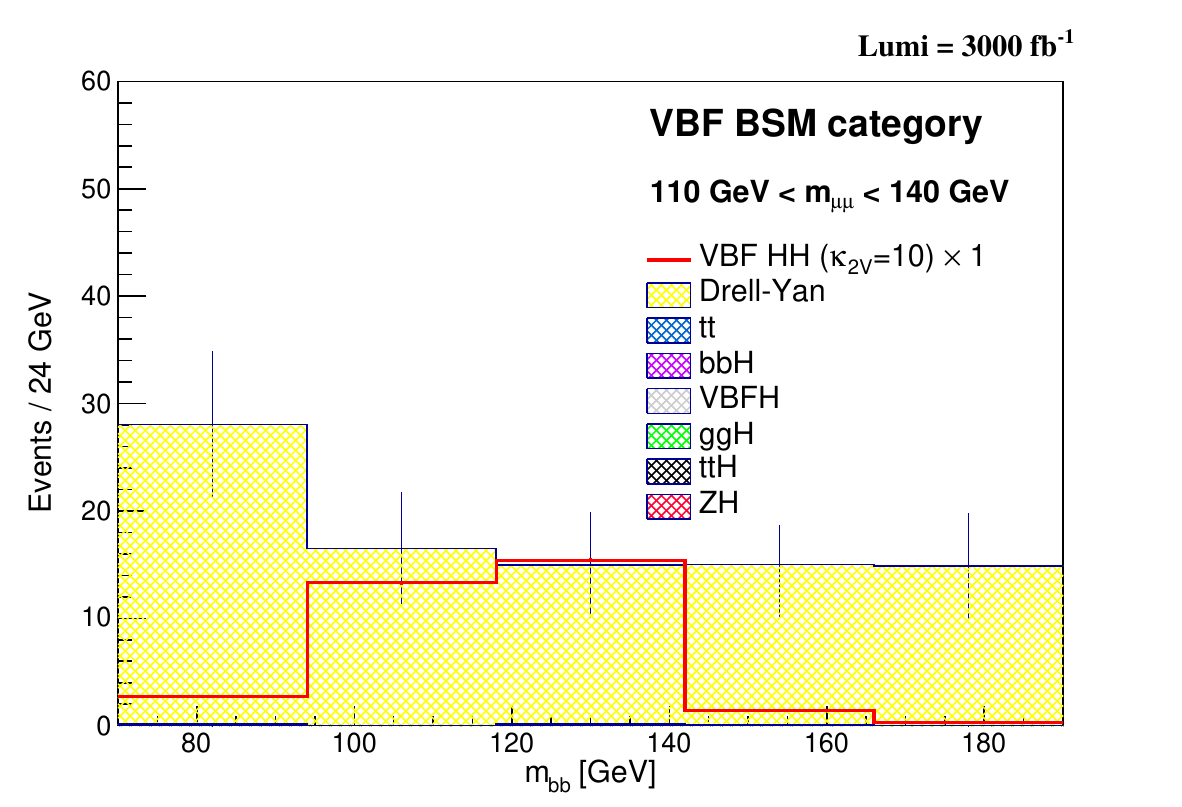}}
   \caption{The distributions of $m_{\mu\mu}$ and $m_{bb}$ in the VBF BSM category from the cut-based analysis.}\label{fig:VBFBSM_2mu2bj}
\end{figure*}

\begin{figure*}[hbt!]
   \centering
   \subcaptionbox{The distribution of $m_{\mu\mu}$ in the VBF SM category.\label{fig:m2mu_VBFSM}}
     {\includegraphics[width=0.45\textwidth]{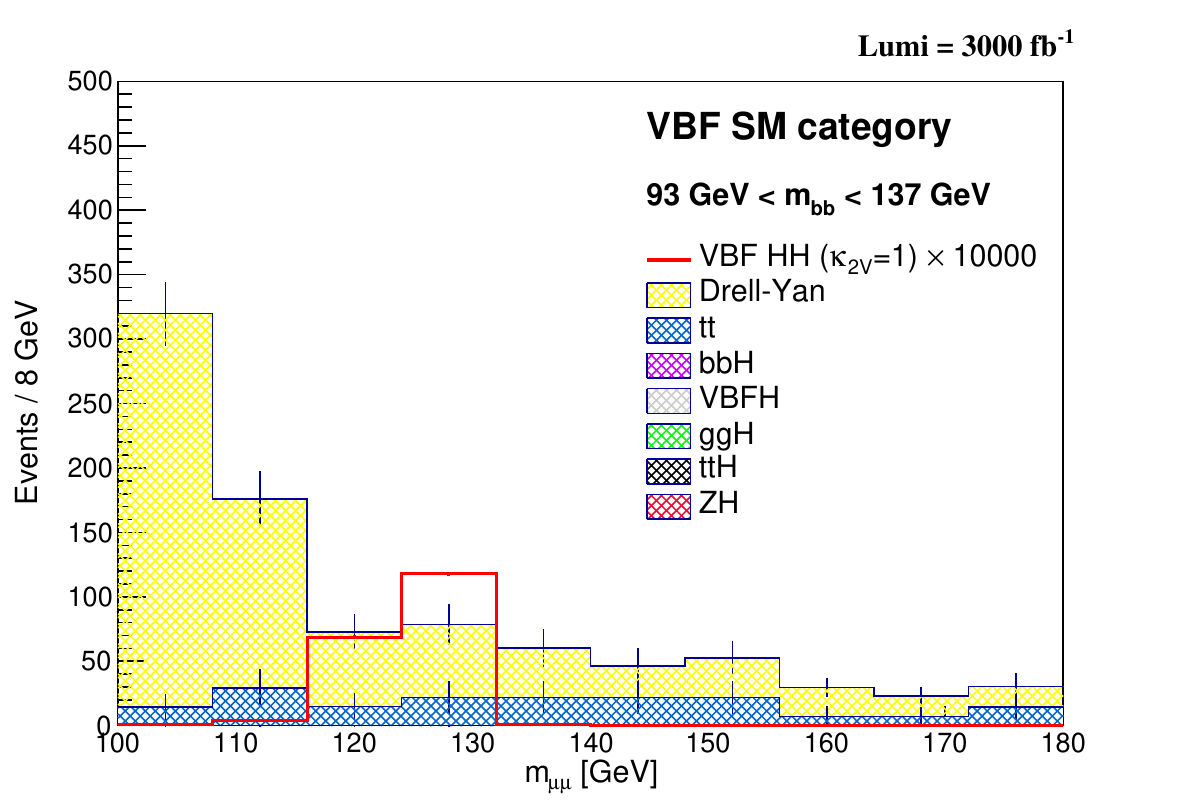}}
   \subcaptionbox{The distribution of $m_{bb}$ in the VBF SM category.\label{fig:m2bj_VBFSM}}
     {\includegraphics[width=0.45\textwidth]{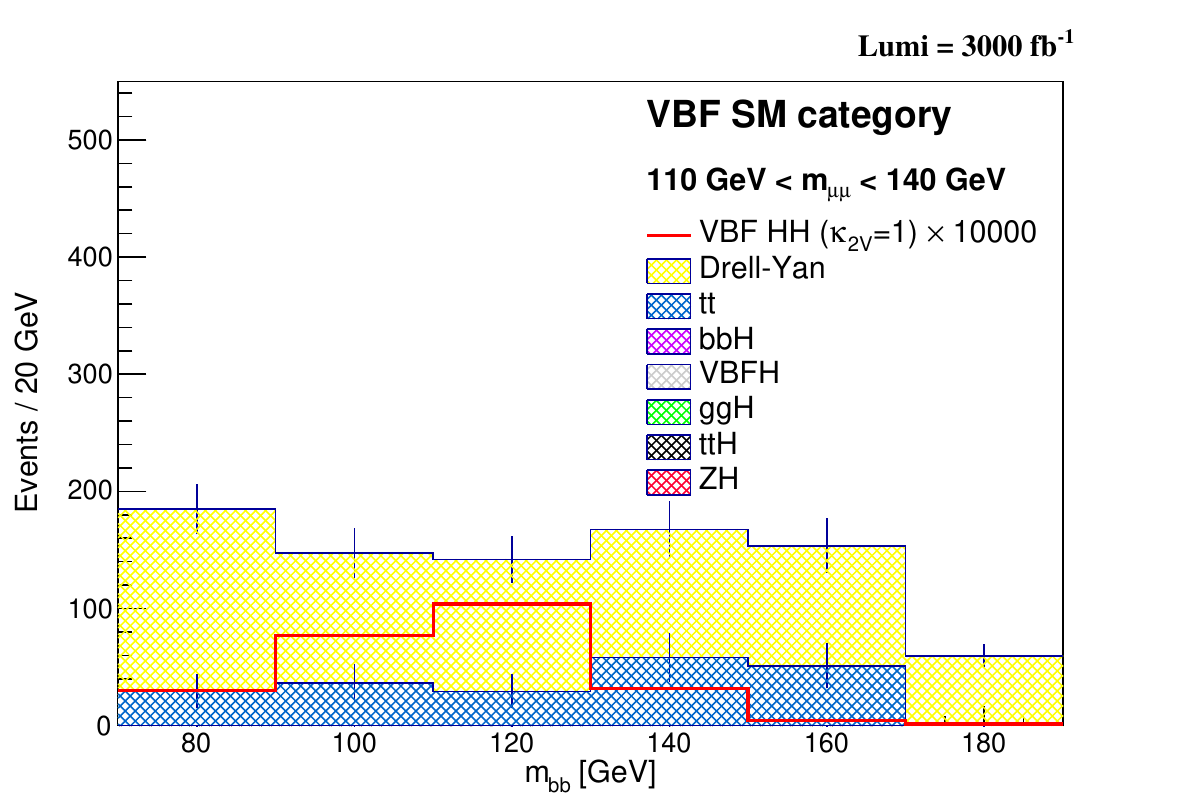}}
   \caption{The distributions of $m_{\mu\mu}$ and $m_{bb}$ in the VBF SM category from the cut-based analysis.}\label{fig:VBFSM_2mu2bj}
\end{figure*}

\begin{figure*}[hbt!]
   \centering
   \subcaptionbox{The distribution of $m_{\mu\mu}$ in the ggF SM category.\label{fig:sm_2mu}}
     {\includegraphics[width=0.45\textwidth]{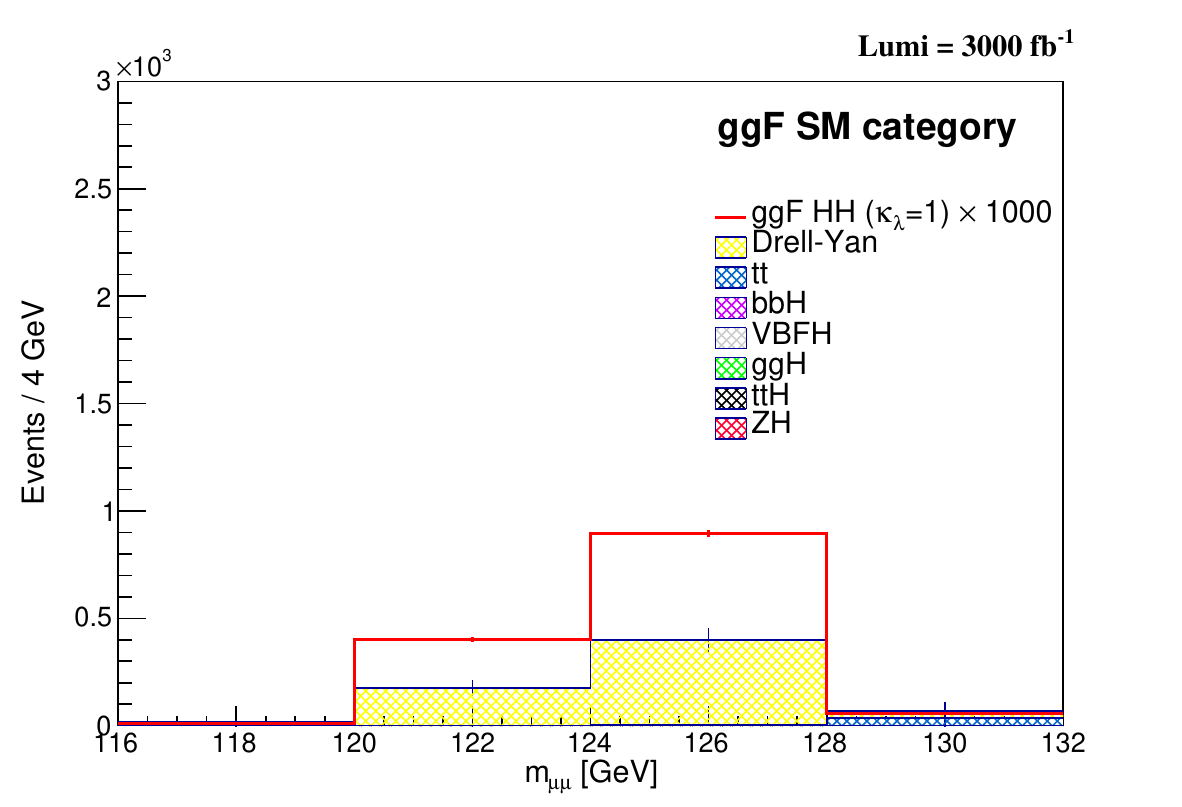}}
   \subcaptionbox{The distribution of $m_{bb}$ in the ggF SM category.\label{fig:sm_2bj}}
     {\includegraphics[width=0.45\textwidth]{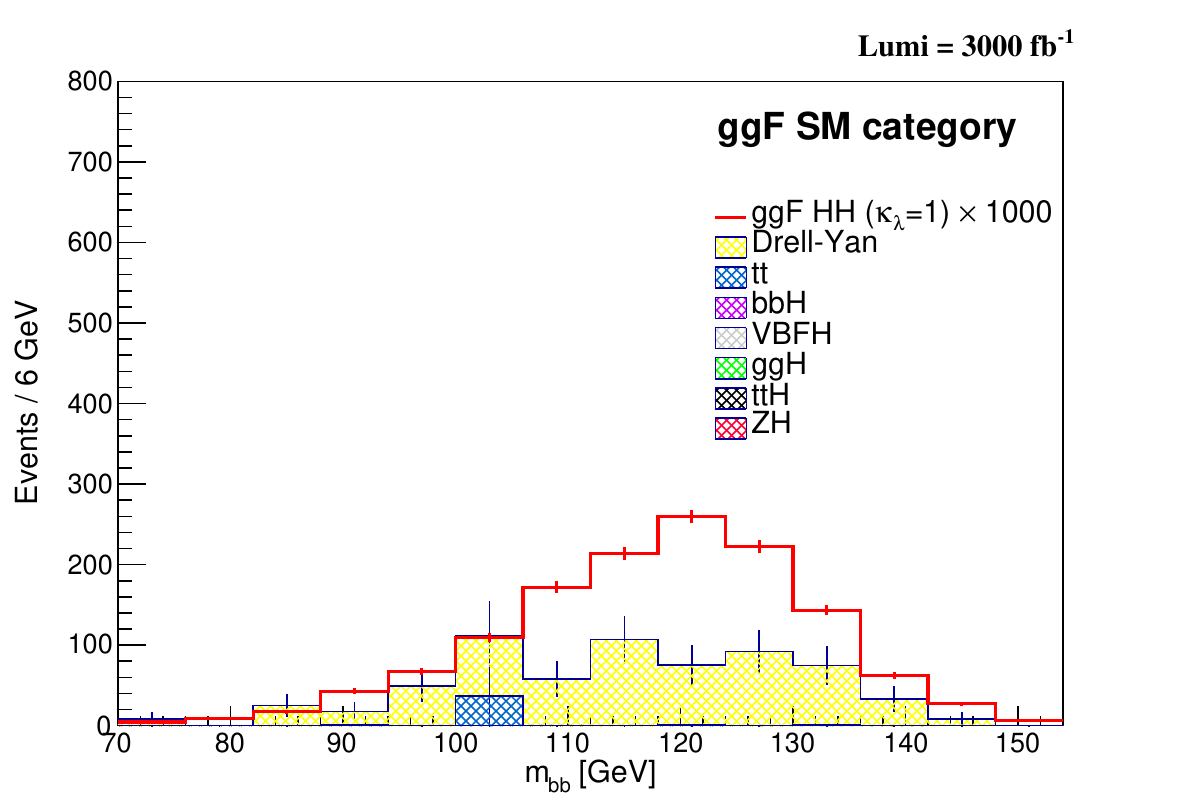}}
   \caption{The distributions of $m_{\mu\mu}$ and $m_{bb}$ in the ggF SM category from the BDT analysis.}\label{fig:sm_2mu2bj}
\end{figure*}

\begin{figure*}[hbt!]
   \centering
   \subcaptionbox{The distribution of $m_{\mu\mu}$ in the ggF BSM category.\label{fig:bsm_2mu}}
     {\includegraphics[width=0.45\textwidth]{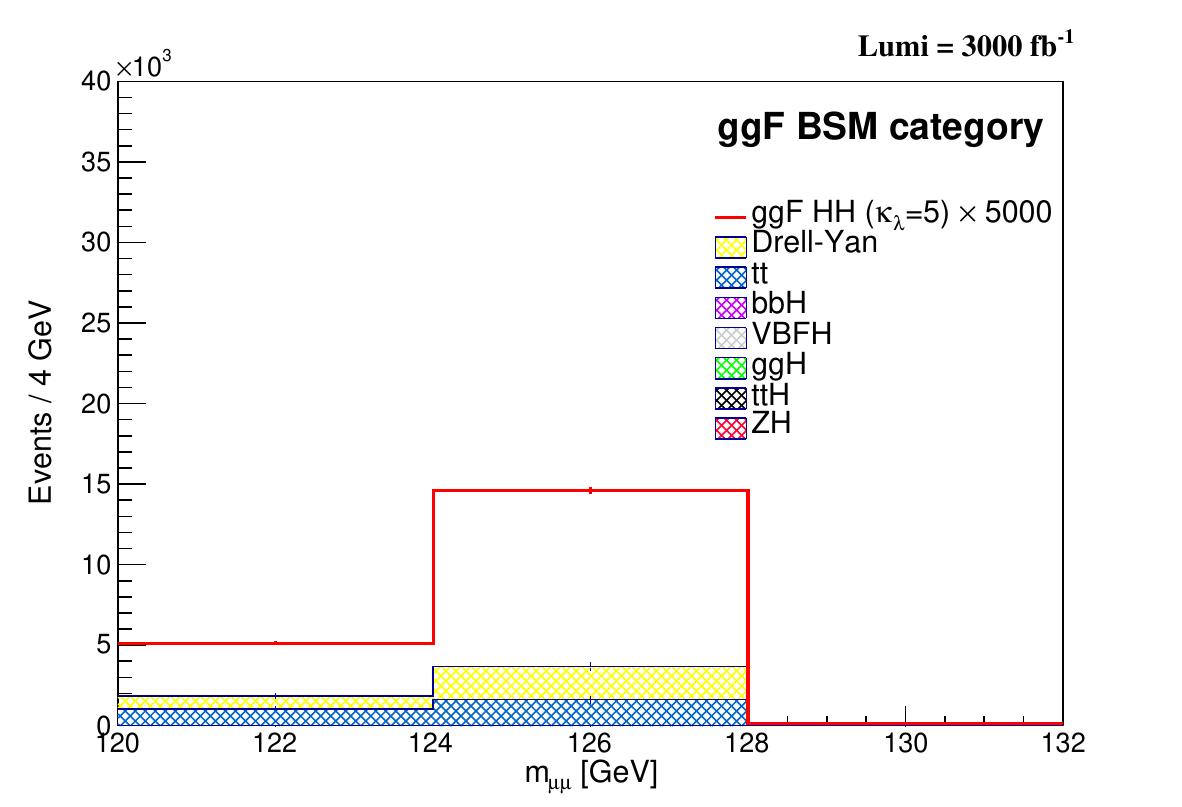}}
   \subcaptionbox{The distribution of $m_{bb}$ in the ggF BSM category.\label{fig:bsm_2bj}}
     {\includegraphics[width=0.45\textwidth]{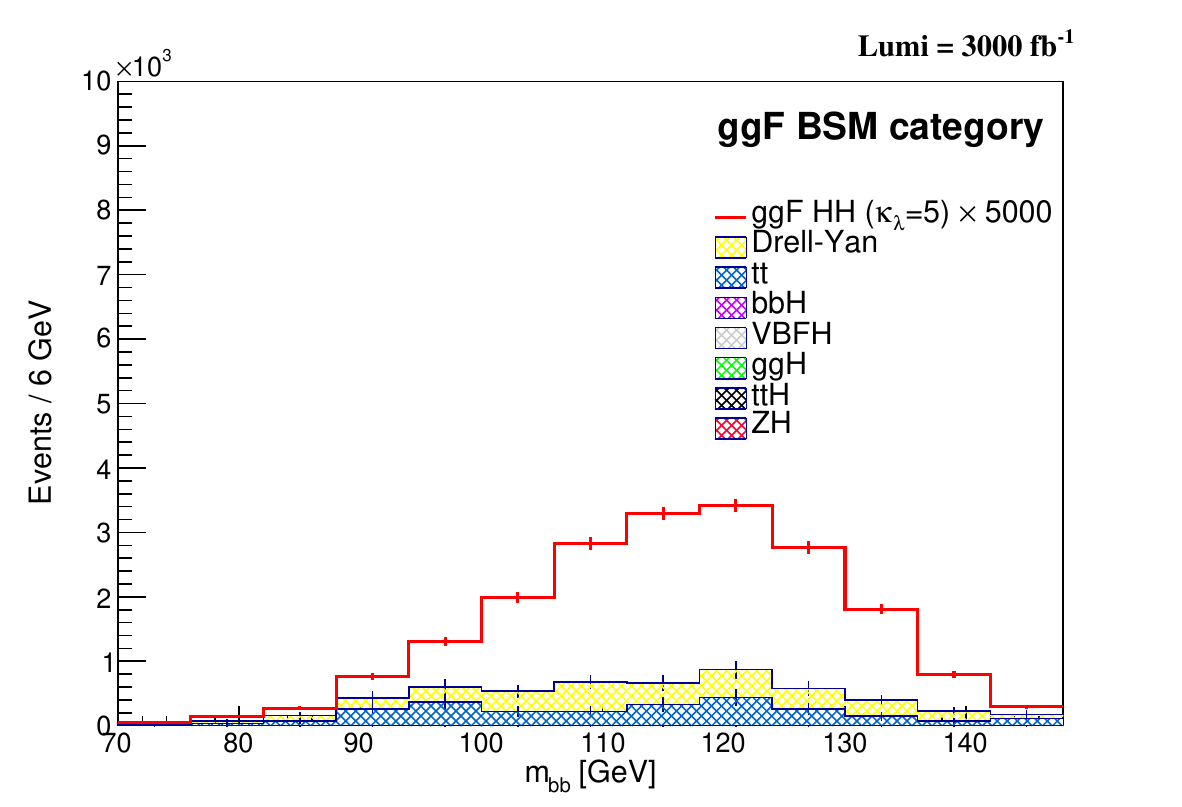}}
   \caption{The distributions of $m_{\mu\mu}$ and $m_{bb}$ in the ggF BSM category from the BDT analysis.}\label{fig:bsm_2mu2bj}
\end{figure*}

All events are used in searching for the signal combining the four categories after the background suppression, using either the cut-based or the BDT approach. The fits are performed in the ranges of $100<m_{\mu\mu}<180\ \rm{GeV}$ and $70<m_{bb}<190\ \rm{GeV}$ in the four categories. The fitting templates are the combined di-muon mass $m_{\mu\mu}$ and di-bjet mass $m_{bb}$ distributions as shown in FIG.~\ref{fig:m2mu_ggFSM} to~\ref{fig:m2bj_VBFSM} for all the four categories with the cut-based method, and in FIG.~\ref{fig:sm_2mu} to~\ref{fig:bsm_2bj} for the two ggF categories with the BDT approach. 


With a integrated luminosity of 3000 fb$^{-1}$, the search with $\text{HH}\to b\bar{b}\mu^+\mu^-$ does not reach the threshold for a discovery. The upper limits on the cross-sections at 95\% confidence level (CL) are extracted using the modified frequentist CL$_{s}$ approach~\cite{Junk:1999kv,Read:2002hq} in which the asymptotic approximation~\cite{Cowan:2010js} is applied. For the SM ggF and VBF HH cross-section, their upper limits at 95\% CL are presented in the unit of their SM cross-section as shown in TAB.~\ref{table:4} for the integrated luminosities of 300 fb$^{-1}$ from the LHC Run3 only, 450 fb$^{-1}$ from the combined Run2 and Run3, and 3000 fb$^{-1}$ from the full runs in the LHC and the HL-LHC. The expected upper limit at 95\% CL on the ggF HH prediction is 47 (28) times of its SM prediction using the cut-based (BDT) approach with the full   integrated luminosity. The expected upper limit at 95\% CL on the VBF HH prediction is 928 times of its SM prediction using the cut-based approach with the full integrated luminosity. The expected upper limits are also set for the combined ggF and VBF HH processes assuming their SM cross-sections. Given the small VBF rate, its contribution to the upper limits of the combined ggF and VBF HH is marginal. All results are extracted assuming $m_{\rm{H}}=125\ \rm{GeV}$ that is close to the most precise measurement of the Higgs boson mass to date $m_{\rm{H}}=125.38\pm0.14\ \rm{GeV}$~\cite{CMS:2020xrn}.

\begin{table}[hbt!]
\renewcommand{\arraystretch}{1.5}
\centering
\begin{tabular}{ c c c c } 
\hline
\hline
Analysis Type & 300 fb$^{-1}$ & 450 fb$^{-1}$ & 3000 fb$^{-1}$ \\
\hline
 & \multicolumn{3}{c}{ggF $\rm{HH}$ ($\sigma/\sigma_{\rm{SM}}$)} \\
 \hline
Cut-based & $152^{+87}_{-46}$ & $123^{+70}_{-37}$ & $47^{+26.1}_{-14.1}$ \\ 
BDT & $96^{+56}_{-29.8}$ & $77^{+45}_{-23.9}$ & $28^{+16.3}_{-8.8}$ \\
 \hline
  & \multicolumn{3}{c}{ggF+VBF $\rm{HH}$ ($\sigma/\sigma_{\rm{SM(ggF+VBF)}}$)} \\
 \hline
Cut-based & $152^{+86}_{-46}$ & $122^{+70}_{-37}$ & $46^{+26.1}_{-13.9}$ \\ 
BDT & $96^{+56}_{-29.7}$ & $77^{+45}_{-23.9}$ & $28^{+16.2}_{-8.8}$ \\
 \hline
 & \multicolumn{3}{c}{VBF $\rm{HH}$ ($\sigma/\sigma_{\rm{SM}}$)} \\
\hline
 Cut-based & $3195^{+1440}_{-960}$ & $2555^{+1130}_{-760}$ & $928^{+380}_{-265}$ \\
\hline
\hline
\end{tabular}
\caption{The upper limits at 95\% CL of the ggF and VBF HH cross-section in the cut-based and BDT analyses.}
\label{table:4}
\end{table}

\begin{figure*}[hbt!]
   \centering
   \subcaptionbox{Expected 95\% CL upper limits of the ggF HH cross-section as a function of $\kappa_{\lambda}$ in the individual categories (Cat.) and the combination.\label{fig:muti-kl}}
     {\includegraphics[width=0.45\textwidth]{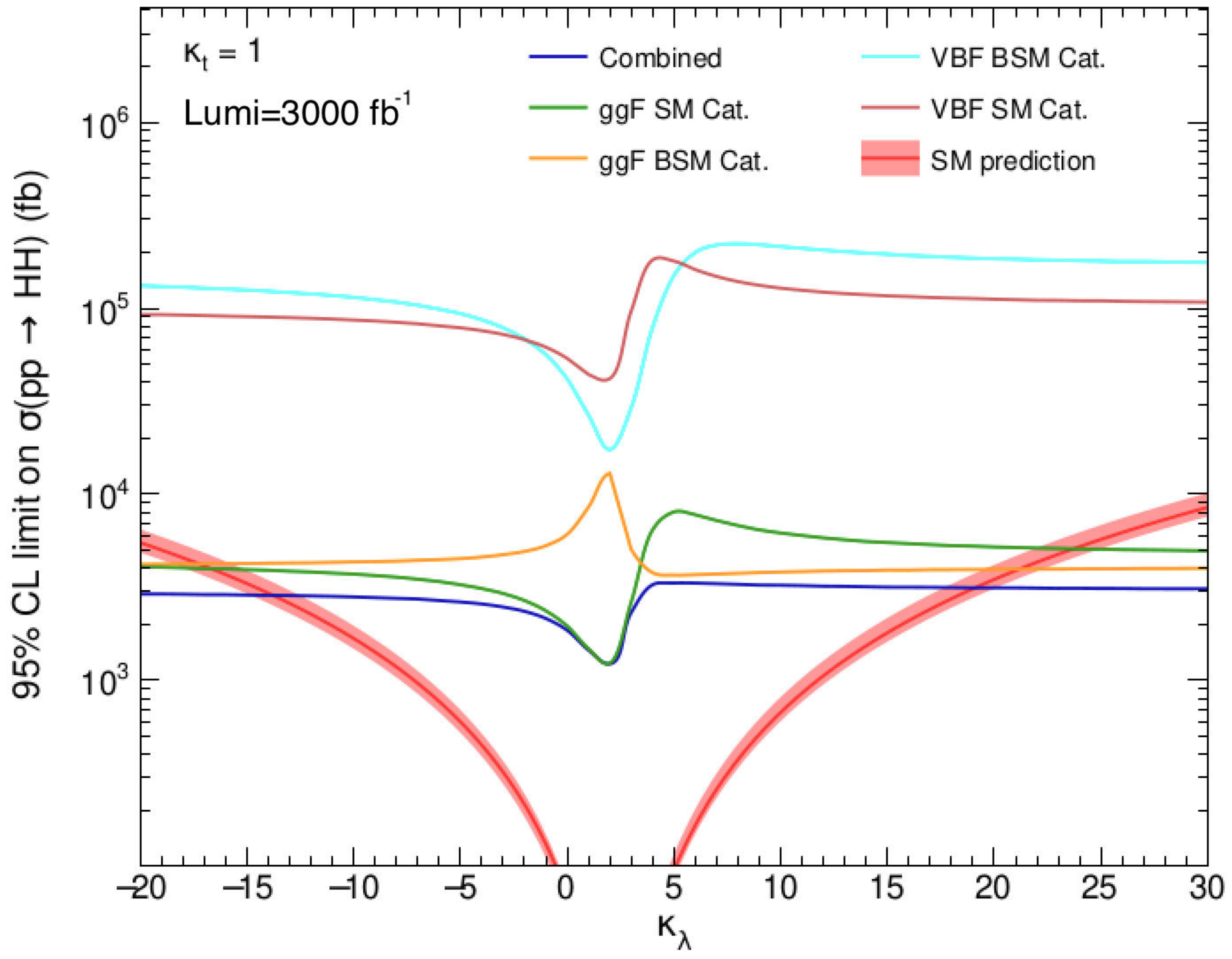}}
   \subcaptionbox{Expected 95\% CL upper limits of the ggF HH cross-section as a function of $\kappa_{\lambda}$.\label{fig:kl}}
     {\includegraphics[width=0.45\textwidth]{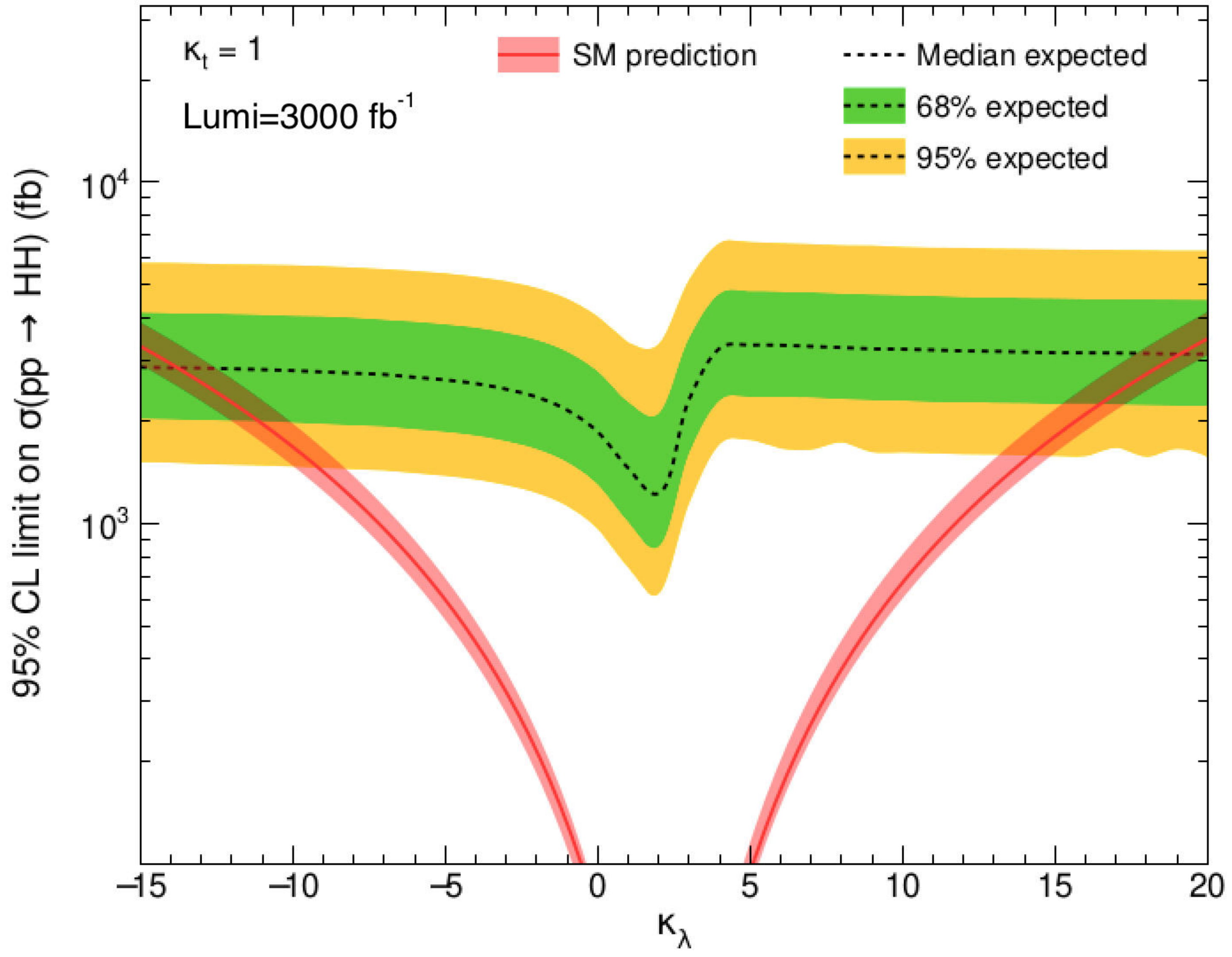}}
   \caption{Expected 95\% CL upper limits of the ggF HH cross-section for the breakdown and the combination as a function of $\kappa_{\lambda}$, using the cut-based approach. The green and yellow bands surrounding the upper limit median represent its 68\% and 95\% CL uncertainty. The red solid curve with its band represents the theoretical prediction and the corresponding uncertainty~\cite{LHC-HH}.}\label{fig:kl_muti-kl}
\end{figure*}

\begin{figure*}[hbt!]
   \centering
   \subcaptionbox{Expected 95\% CL upper limits of the ggF HH cross-section as a function of $\kappa_{\lambda}$ in the individual categories (Cat.) and the combination.\label{fig:xgb-muti-kl}}
     {\includegraphics[width=0.45\textwidth]{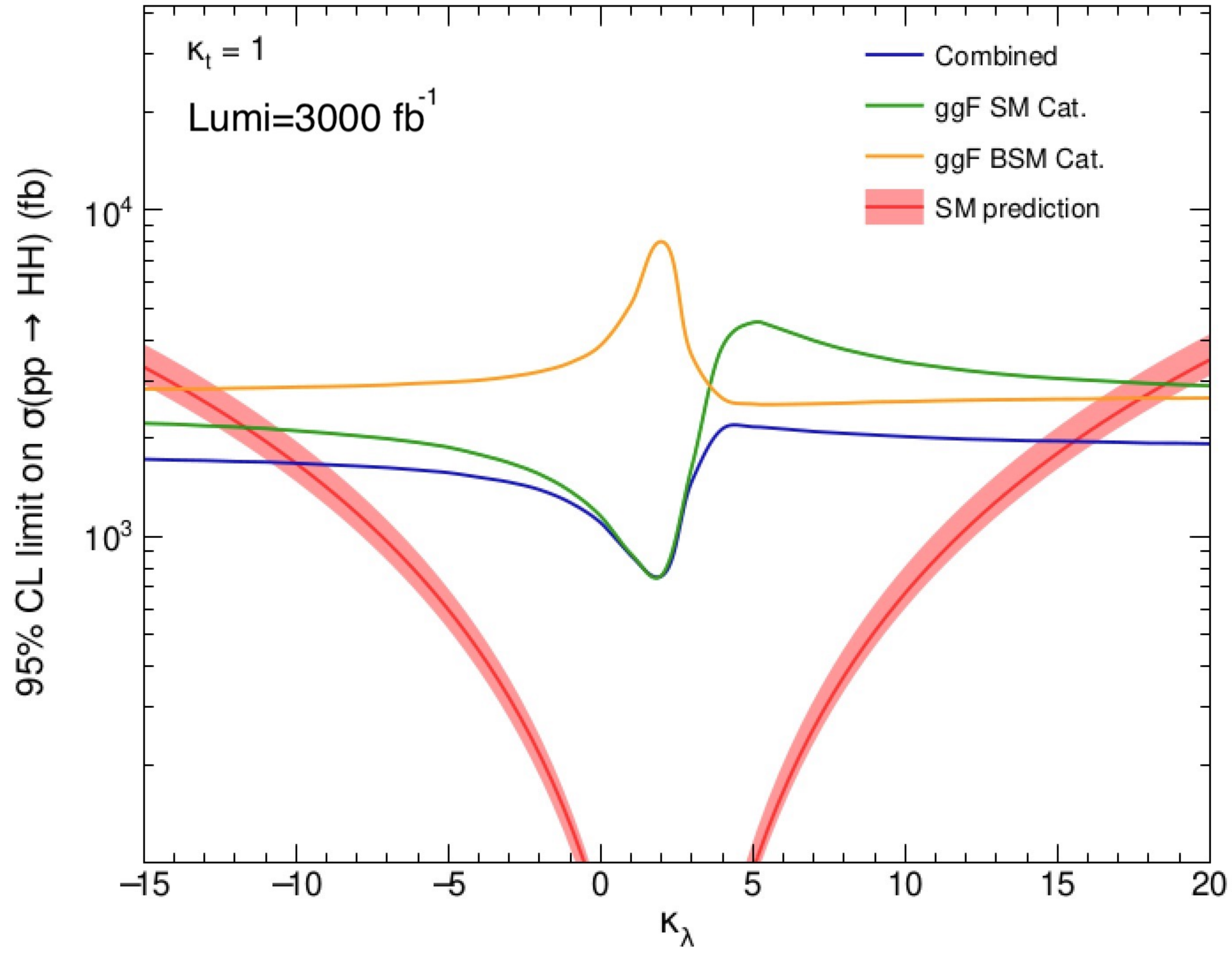}}
   \subcaptionbox{Expected 95\% CL upper limits of the ggF HH cross-section as a function of $\kappa_{\lambda}$.\label{fig:xgb-kl}}
     {\includegraphics[width=0.45\textwidth]{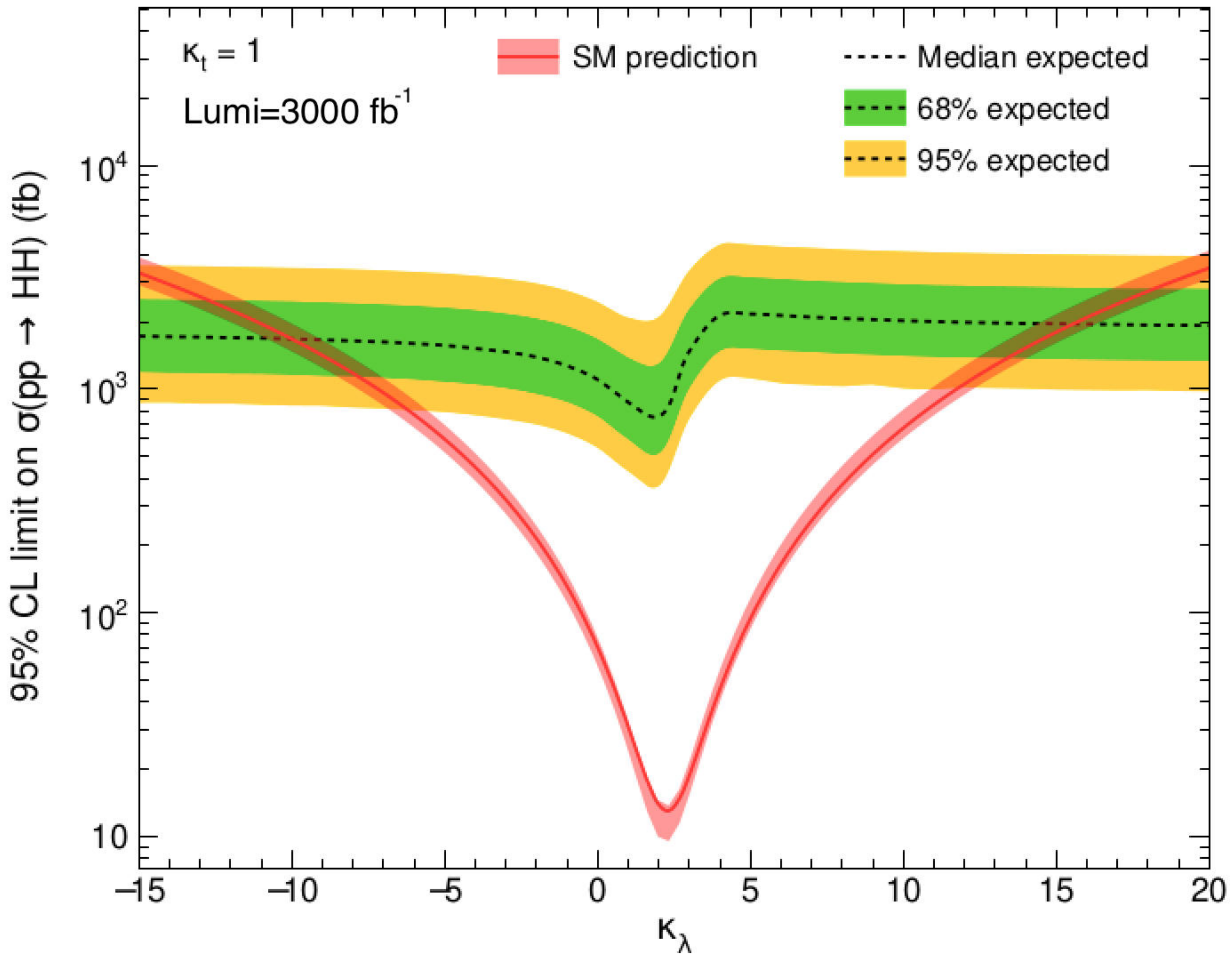}}
   \caption{Expected 95\% CL upper limits of the ggF HH cross-section for the breakdown and the combination as a function of $\kappa_{\lambda}$ using the BDT approach. The green and yellow bands surrounding the upper limit median represent its 68\% and 95\% CL uncertainty. The red solid curve with its band represents the theoretical prediction and the corresponding uncertainty~\cite{LHC-HH}.}\label{fig:xgb_kl_muti-kl}
\end{figure*}

Benefiting from the four categories that are defined to maximize the sensitivity not only at the SM coupling but also to the BSM, the upper limits at 95\% CL are scanned along $\kappa_\lambda$ and $\kappa_{\text{2V}}$, assuming the top quark Yukawa coupling and the HVV coupling SM-like ($\kappa_{\rm{t}}=1$ and $\kappa_{\text{V}}$=1). The scan on $\kappa_\lambda$ is shown in FIG.~\ref{fig:muti-kl} with the individual contributions from each category and the combination, and in FIG.~\ref{fig:kl} with the combined results, using the cut-based approach. The green and yellow bands surrounding the upper limit median represent its 68\% and 95\% CL uncertainty. The red solid curve with its band represents the theoretical prediction and the corresponding uncertainty~\cite{LHC-HH}. At 95\% CL, the expected constraint on $\kappa_{\lambda}$ is $-13.8 < \kappa_{\lambda}< 19.1$ with the full integrated luminosity. The same scan is performed using the BDT approach as shown in FIG.~\ref{fig:xgb-muti-kl} for the breakdown and FIG.~\ref{fig:xgb-kl} for the combination. At 95\% CL, the expected constraint on $\kappa_{\lambda}$ using the BDT approach is $-10.0 < \kappa_{\lambda}< 15.5$ with the full integrated luminosity.


\begin{figure*}[hbt!]
   \centering
   \subcaptionbox{Expected 95\% CL upper limits of the VBF HH cross-section as a function of $\kappa_{\text{2V}}$ in the individual categories (Cat.) and the combination.\label{fig:muti-C2V}}
     {\includegraphics[width=0.45\textwidth]{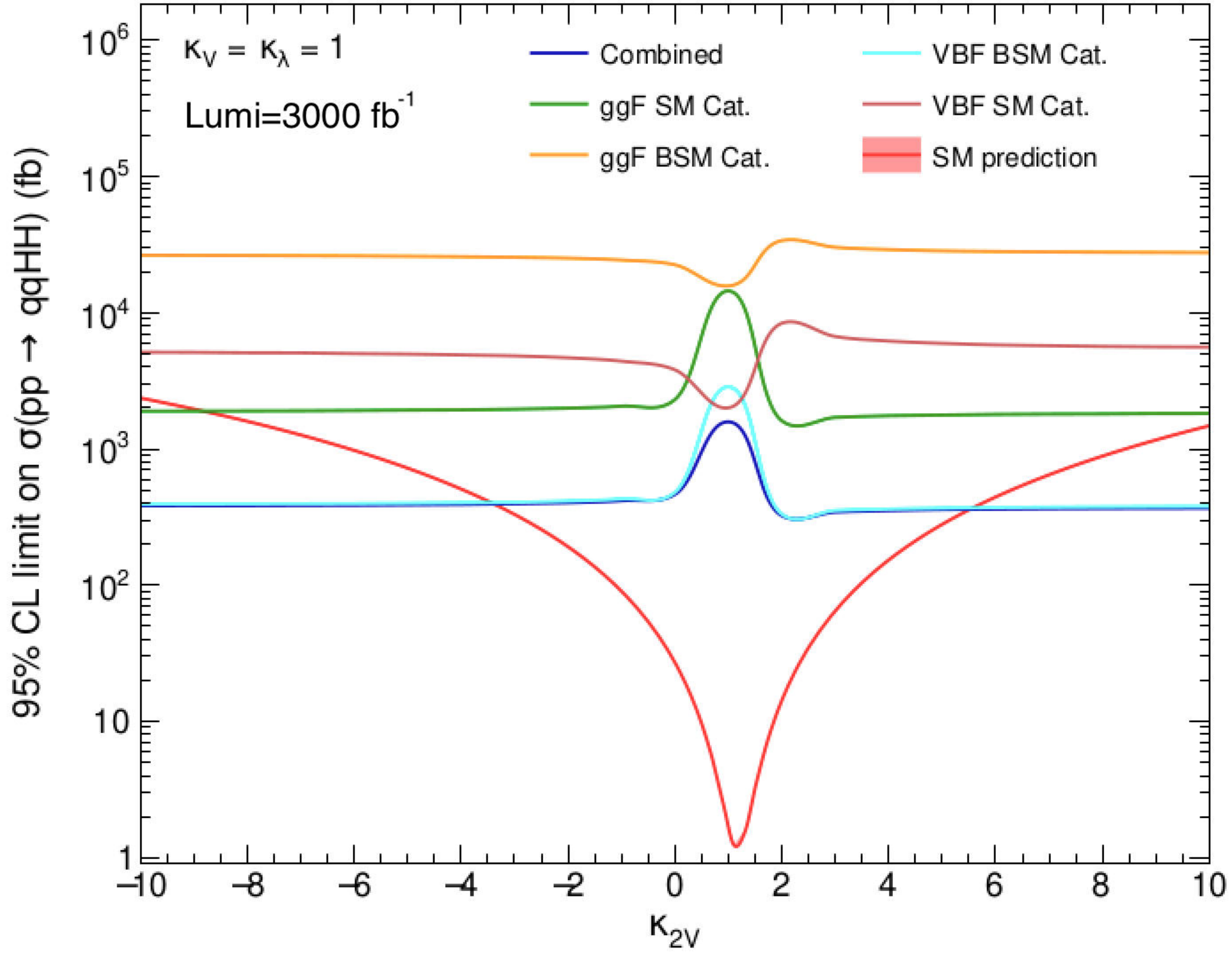}}
   \subcaptionbox{Expected 95\% CL upper limits of the VBF HH cross-section as a function of $\kappa_{\text{2V}}$.\label{fig:C2V}}
     {\includegraphics[width=0.45\textwidth]{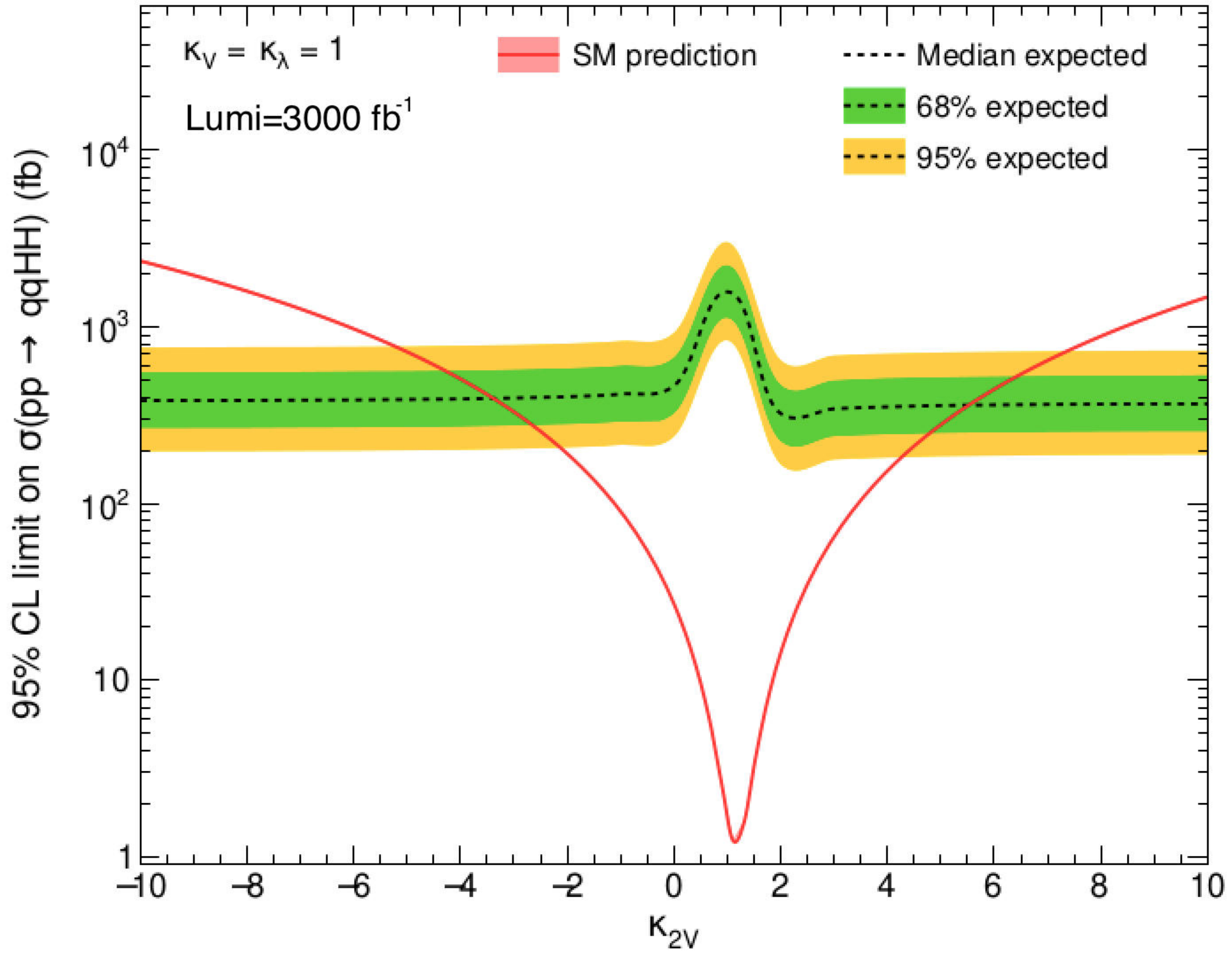}}
   \caption{Expected 95\% CL upper limits of the VBF HH cross-section for the breakdown and the combination as a function of $\kappa_{\text{2V}}$ using the cut-based approach. The green and yellow bands surrounding the upper limit median represent its 68\% and 95\% CL uncertainty. The red solid curve with its band represents the theoretical prediction and the corresponding uncertainty~\cite{LHC-HH}.}\label{fig:C2V_muti-C2V}
\end{figure*}

Similarly, the cross-section upper limits at 95\% CL are also scanned on $\kappa_{\text{2V}}$ as shown in FIG.~\ref{fig:muti-C2V} for the individual contributions per category, and FIG.~\ref{fig:C2V} for the combined results with the uncertainty bands. The green and yellow bands surrounding the upper limit median represent its 68\% and 95\% CL uncertainty. The red solid curve with its band represents the theoretical prediction and the corresponding uncertainty~\cite{LHC-HH}. At 95\% CL, the expected constraint on $\kappa_{\rm{2V}}$ is $-3.4 < \kappa_{\rm{2V}}< 5.5$ with the full integrated luminosity.

In the end, the constraints of $\kappa_{\lambda}$ and $\kappa_{\rm{2V}}$ are also extracted for lower integrated luminosities, e.g. 300 fb$^{-1}$ from the LHC Run3 only and 450 fb$^{-1}$ from the combined Run2 and Run3 in TAB.~\ref{table:scan}.


\begin{table}[hbt!]
\renewcommand{\arraystretch}{1.5}
\centering
\begin{tabular}{ c c c c } 
\hline
\hline
Analysis Type & 300 fb$^{-1}$ & 450 fb$^{-1}$ & 3000 fb$^{-1}$ \\
\hline
 & \multicolumn{3}{c}{$\kappa_\lambda$ constraints} \\
 \hline
Cut-based & (-26.9, 32.2) & (-24.0, 29.3) & (-13.8, 19.1) \\ 
BDT & (-20.7, 26.2) & (-18.3, 23.8) & (-10.0, 15.5) \\
 \hline
 & \multicolumn{3}{c}{$\kappa_{\rm{2V}}$ constraints} \\
\hline
 Cut-based & (-7.6, 9.8) & (-6.6, 8.8) & (-3.4, 5.5) \\
\hline
\hline
\end{tabular}
\caption{The 95\% CL constraints on $\kappa_\lambda$ and $\kappa_{\rm{2V}}$ in the cut-based and BDT analyses.}
\label{table:scan}
\end{table}

\section{Conclusions}
\label{sec:sum}

This paper presents a comprehensive study of the Higgs boson pair production in the rare decay of $\text{HH} \to b\bar{b}\mu^+\mu^-$ with both the ggF and VBF production modes included to probe the Higgs self-coupling $\kappa_\lambda$ and the four-boson HHVV coupling $\kappa_{\text{2V}}$ for the first time. As both of the production rates and the kinematics strongly depend on the two couplings of interests, the analysis is performed with four different event categories, each of which focuses on one of the following cases: SM-like $\kappa_{\lambda}$, BSM $\kappa_{\lambda}$, SM-like $\kappa_{\rm{2V}}$ and BSM $\kappa_{\rm{2V}}$. In each of the categories, the corresponding background contributions are suppressed by the dedicated cut-based or BDT approach that is optimized individually. The final result is extracted with fits to the combined spectrum of the di-muon invariant mass $m_{\mu\mu}$ and the di-bjet invariant mass $m_{bb}$. With a integrated luminosity up to 3000 fb$^{-1}$, the channel $\text{HH}\to b\bar{b}\mu^+\mu^-$ can not lead to the observation of HH with the cut-based or the BDT approach discussed in this paper. The upper limits at 95\% confidence level on the cross-sections are then extracted using the full HL-LHC integrated luminosity of 3000 fb$^{-1}$. They are 47 (28) for the ggF HH and 928 for the VBF HH using the cut-based (BDT) approach, both in the unit of their SM predicted cross-section. The cross-section limits are also used to constrain the couplings. The constraints at 95\% confidence level are $-13.8 < \kappa_{\lambda}< 19.1$ ($-10.0 < \kappa_{\lambda}< 15.5$) and $-3.4 < \kappa_{\rm{2V}}< 5.5$ using the cut-based (BDT) approach. 

The recent experimental results of $\text{HH} \to b\bar{b}\gamma\gamma$ and $\text{HH} \to b\bar{b}\tau^+\tau^-$ that lead the HH sensitivity are projected to the HL-LHC with an integrated luminosity of 3000 fb$^{-1}$ under $\sqrt{s} =$ 14 TeV~\cite{ATLAS:2022okt}. To have a fair comparison with our results, the projections with only statistical uncertainties are quoted in the following. The expected constraints on $\kappa_\lambda$ at 95\% confidence level are $1.2 < \kappa_{\lambda}< 4.2$ from $\text{HH} \to b\bar{b}\gamma\gamma$ and $2.4 < \kappa_{\lambda}< 4.5$ from $\text{HH} \to b\bar{b}\tau^+\tau^-$. The expected upper limits on the cross section at 95\% confidence level are 0.86 and 0.49 times the SM prediction for $\text{HH} \to b\bar{b}\gamma\gamma$ and $\text{HH} \to b\bar{b}\tau^+\tau^-$, respectively.




In conclusion, the $\text{HH} \to b\bar{b}\mu^+\mu^-$ decay channel could not lead to the observation alone up to the HL-LHC using the method discussed in this paper, but it is still able to contribute in a sizeable way to the HH search combination and can be sensitive to BSM enhancement given its small rate and excellent di-muon mass peak.


\begin{acknowledgments}

We thank Qiang Li, Congqiao Li and Sitian Qian for the helpful discussions. The work is supported in part by the National Science Foundation of China under Grants No. 12175006, No. 12188102 and No. 12061141002, and by Peking University under startup Grant No. 7100603613.

\end{acknowledgments}

\nocite{*}


\bibliography{apssamp}

\end{document}